\documentclass[fleqn,usenatbib]{mnras}

\usepackage{newtxtext,newtxmath}
\usepackage{multirow}

\usepackage[utf8]{inputenc}
\usepackage[T1]{fontenc}
\usepackage{ae,aecompl}

\usepackage{graphicx}	
\usepackage{amsmath}	
\usepackage{amssymb}	
\usepackage{enumerate}

\usepackage[left]{lineno}

\title[Measurement of the EBL with MAGIC and \lat]{Measurement of the Extragalactic Background Light using MAGIC and \lat\ gamma-ray observations of blazars up to z = 1}

\author[MAGIC Collaboration]
{\parbox{\textwidth}{
MAGIC Collaboration: 
V.~A.~Acciari$^{1}$,
S.~Ansoldi$^{2}$,
L.~A.~Antonelli$^{3}$,
A.~Arbet Engels$^{4}$,
D.~Baack$^{5}$,
A.~Babi\'c$^{6}$,
B.~Banerjee$^{7}$,
U.~Barres de Almeida$^{8}$,
J.~A.~Barrio$^{9}$,
J.~Becerra Gonz\'alez$^{1}$
% ...
}
\\
\newauthor
{\em \large \raggedright  (Affiliations can be found after the references)}
}
\newcommand{\writeaffiliations}{
\section*{\centering Affiliations}
$^{1}$ {Inst. de Astrof\'isica de Canarias, E-38200 La Laguna, and Universidad de La Laguna, Dpto. Astrof\'isica, E-38206 La Laguna, Tenerife, Spain} \\
$^{2}$ {Universit\`a di Udine, and INFN Trieste, I-33100 Udine, Italy} \\
$^{3}$ {National Institute for Astrophysics (INAF), I-00136 Rome, Italy} \\
$^{4}$ {ETH Zurich, CH-8093 Zurich, Switzerland} \\
$^{5}$ {Technische Universit\"at Dortmund, D-44221 Dortmund, Germany} \\
$^{6}$ {Croatian Consortium: University of Rijeka, Department of Physics, 51000 Rijeka; University of Split - FESB, 21000 Split; University of Zagreb - FER, 10000 Zagreb; University of Osijek, 31000 Osijek; Rudjer Boskovic Institute, 10000 Zagreb, Croatia} \\
$^{7}$ {Saha Institute of Nuclear Physics, HBNI, 1/AF Bidhannagar, Salt Lake, Sector-1, Kolkata 700064, India} \\
$^{8}$ {Centro Brasileiro de Pesquisas F\'isicas (CBPF), 22290-180 URCA, Rio de Janeiro (RJ), Brasil} \\
$^{9}$ {Unidad de Part\'iculas y Cosmolog\'ia (UPARCOS), Universidad Complutense, E-28040 Madrid, Spain} \\
$^{10}$ {University of \L\'od\'z, Department of Astrophysics, PL-90236 \L\'od\'z, Poland} \\
$^{11}$ {Universit\`a  di Siena and INFN Pisa, I-53100 Siena, Italy} \\
$^{12}$ {Deutsches Elektronen-Synchrotron (DESY), D-15738 Zeuthen, Germany} \\
$^{13}$ {Istituto Nazionale Fisica Nucleare (INFN), 00044 Frascati (Roma) Italy} \\
$^{14}$ {Max-Planck-Institut f\"ur Physik, D-80805 M\"unchen, Germany} \\
$^{15}$ {Institut de F\'isica d'Altes Energies (IFAE), The Barcelona Institute of Science and Technology (BIST), E-08193 Bellaterra (Barcelona), Spain} \\
$^{16}$ {Universit\`a di Padova and INFN, I-35131 Padova, Italy} \\
$^{17}$ {Universit\`a di Pisa, and INFN Pisa, I-56126 Pisa, Italy} \\
$^{18}$ {Universit\"at W\"urzburg, D-97074 W\"urzburg, Germany} \\
$^{19}$ {Finnish MAGIC Consortium: Tuorla Observatory (Department of Physics and Astronomy) and Finnish Centre of Astronomy with ESO (FINCA), University of Turku, FI-20014 Turku, Finland; Astronomy Division, University of Oulu, FI-90014 Oulu, Finland} \\
$^{20}$ {Departament de F\'isica, and CERES-IEEC, Universitat Aut\`onoma de Barcelona, E-08193 Bellaterra, Spain} \\
$^{21}$ {Universitat de Barcelona, ICCUB, IEEC-UB, E-08028 Barcelona, Spain} \\
$^{22}$ {ICRANet-Armenia at NAS RA, 0019 Yerevan, Armenia} \\
$^{23}$ {Japanese MAGIC Consortium: ICRR, The University of Tokyo, 277-8582 Chiba, Japan; Department of Physics, Kyoto University, 606-8502 Kyoto, Japan; Tokai University, 259-1292 Kanagawa, Japan; RIKEN, 351-0198 Saitama, Japan} \\
$^{24}$ {Inst. for Nucl. Research and Nucl. Energy, Bulgarian Academy of Sciences, BG-1784 Sofia, Bulgaria} \\
$^{25}$ {Humboldt University of Berlin, Institut f\"ur Physik D-12489 Berlin Germany} \\
$^{26}$ {also at Dipartimento di Fisica, Universit\`a di Trieste, I-34127 Trieste, Italy} \\
$^{27}$ {also at Port d'Informaci\'o Cient\'ifica (PIC) E-08193 Bellaterra (Barcelona) Spain} \\
$^{28}$ {also at INAF-Trieste and Dept. of Physics \& Astronomy, University of Bologna, Italy}
} 

\date{Accepted XXX. Received YYY; in original form ZZZ}

\pubyear{2018}

\def\eg{{e.g.~}}
\def\lat{{{\em Fermi}-LAT}}

\begin{document}
\label{firstpage}
\pagerange{\pageref{firstpage}--\pageref{lastpage}}
\maketitle

\clearpage

\begin{abstract}
We present a measurement of the extragalactic background light (EBL) based on a joint likelihood analysis of 32 gamma-ray spectra for 12 blazars in the redshift range $z=0.03-0.944$, obtained by the MAGIC telescopes and \lat. The EBL is the part of the diffuse extragalactic radiation spanning the ultraviolet, visible and infrared bands. Major contributors to the EBL are the light emitted by stars through the history of the universe, and the fraction of it which was absorbed by dust in galaxies and re-emitted at longer wavelengths. The EBL can be studied indirectly through its effect on very-high energy photons that are emitted by cosmic sources and absorbed via $\gamma\gamma$ interactions during their propagation across cosmological distances. We obtain estimates of the EBL density in good agreement with state-of-the-art models of the EBL production and evolution. The $1\sigma$ upper bounds, including systematic uncertainties, are between 13\% and 23\% above the nominal EBL density in the models. No anomaly in the expected transparency of the universe to gamma rays is observed in any range of optical depth. We also perform a wavelength-resolved EBL determination, which results in a hint of an excess of EBL in the 0.18 - 0.62 ${\rm \mu m}$ range relative to the studied models, yet compatible with them within systematics.

\end{abstract}




\section{Introduction}
The extragalactic background light (EBL) is a cosmic diffuse radiation field that encloses essential information about galaxy evolution and cosmology (see e.g. \citealt{hauser01,dwek13,dominguez13b} and references therein). It is mainly composed of the ultraviolet, optical, and near-infrared light emitted by stars through the history of the universe, possibly including light from the (yet undetected) population-III stars \citep[\eg][]{inoue14}. A fraction of these photons is absorbed by interstellar dust and re-emitted at longer wavelengths, producing the characteristic double peak spectral energy distribution of the EBL. This radiation is accumulated over the cosmic history, and redshifted by the expansion of the universe. There may be additional contributions to the EBL, such as those connected to accretion processes onto super-massive black holes \citep[\eg][]{kormendy13,shankar16}, or even more exotic sources such as products of the decay of relic dark matter particles \citep[\eg][]{murase12}. 
\par
The direct detection of the EBL using absolute photometry is challenging because of strong foregrounds, mainly zodiacal light but also the brightness of our own Galaxy \citep[\eg][]{arendt98,gorjian00}. Therefore, attempts at direct detection are subject to large uncertainties and biases \citep[\eg][]{matsumoto05, bernstein07,matsuoka11,mattila17}. Other methods focus on measuring the background anisotropies, which still provides inconclusive results \citep[\eg][]{helgason14,zemcov14,zemcov17,helgason17,matsuura17}. None of these techniques provides direct information about the evolution of the EBL with cosmic redshift.

\renewcommand{\thefootnote}{\fnsymbol{footnote}}
\footnotetext[1]{Corresponding authors: A. Moralejo (moralejo@ifae.es), M. Nievas Rosillo (mnievas@ucm.es), A. Domínguez (alberto@gae.ucm.es), T. Hassan (thassan@ifae.es), G. Vanzo (gvanzo@iac.es), M. V\'azquez Acosta (monicava@iac.es)}
\renewcommand{\thefootnote}{\arabic{footnote}}

An alternative methodology to estimate the EBL is based on counting photons in different photometric bands using data from deep galaxy surveys \citep[\eg][]{madau00,fazio04,keenan10,tsumura13,driver16}. This procedure results in EBL estimates that can be considered lower limits, since light from faint undetected galaxy populations or from the outer regions of normal galaxies may be missed \citep[\eg][]{bernstein02c}. Furthermore, cosmic variance may contribute to systematic uncertainties using this technique \citep{somerville04}.

Efforts centered on building models of the EBL utilize different complementary strategies. Following the classification by \citet{dominguez11a}, these models are divided in four different classes: (1) Forward evolution models that use semi-analytical models of galaxy formation \citep[\eg][]{somerville12,gilmore12,inoue13}, (2) Backward evolution models based on local or low redshift galaxy data, which are extrapolated to higher redshifts making some assumptions on galaxy evolution \citep[\eg][]{franceschini08,franceschini17}, (3) Inferred evolution from the cosmic star formation history of the universe \citep[\eg][]{kneiske02,finke10,khaire15,andrews18} and (4) Observed evolution based on galaxy data over a broad range of redshift \citep[\eg][]{dominguez11a,helgason12,stecker16}. Basically, these models converge to spectral intensities that are close or even match those derived from galaxy counts, at least around the shorter wavelength peak. Uncertainties are larger at the far-IR peak, since most of these models do not include data at those wavelengths, and the luminosity evolution is much faster and more difficult to trace because of source confusion and other instrument limitations \citep[\eg][]{2000AJ....119.2092B,2001PASJ...53...37T,2010A&A...518L..30B}. 

Another technique that has become rather successful to constrain the EBL is based on the observation of gamma rays from distant extragalactic sources. This strategy relies on the fact that photons with energies larger than about 10 GeV traveling cosmological distances suffer an energy- and distance-dependent attenuation by pair-production interaction with the EBL \citep{nikishov62,gould66}. In general, this technique is based on making more or less sophisticated assumptions on the intrinsic/unattenuated energy spectra of the sources, which allows, by comparison with the observed spectra, to derive information on the EBL and, very importantly, on its evolution. Early attempts provided upper limits on the background intensity \citep[\eg][]{stecker92,aharonian06,mazin07,meyer12}. Yet, more recently, thanks to the availability of more and better gamma-ray data, the EBL detection has been claimed by different groups \citep{ackermann12,abramowski13,dominguez13a,biteau15,ahnen16,HESS2017ebl,FermiEBL2018}. These EBL detections constrain the background intensities to be close to the lower limits provided by galaxy counts (within a factor of two or smaller, depending on the energy). However, they are in strong tension with those intensities obtained from early direct detection attempts such as the one presented by \citet{matsumoto05}, \citet{2015ApJ...807...57M},  and \citet{bernstein07}, yet still compatible, or slightly in tension, with more recent estimates such as those by \citet{matsuoka11}, \citet{matsuura1}, and \citet{mattila17}.

Although great progress has been achieved in the study of the EBL in the past years, more work is definitely necessary, particularly in the study of EBL evolution and its high-redshift properties. Interestingly, in the last years, the MAGIC imaging atmospheric Cherenkov telescopes (IACTs) have detected the two farthest sources to date in the very-high-energy band, both of them at $z\sim 1$ \citep{pks1441,b0218}. These detections significantly expand the redshift range of sources available for gamma-ray attenuation measurements from the ground.

In this paper we present EBL constraints based on a joint likelihood analysis of twelve blazars observed with MAGIC during extensive campaigns totalling over 300 hours of exposure, including observations of the most distant very-high-energy (VHE) sources detected. Additionally, we add lower energy data (from 0.1 to $\simeq$ 100 GeV) taken by the Large Area Telescope (LAT) on-board the {\it Fermi Gamma-ray Space Telescope} during similar time ranges as the MAGIC observations. The combination of contemporaneous MAGIC and LAT data allows us to have a better estimate of the intrinsic spectral energy distribution of a given blazar, since the energy range covered by LAT is only slightly affected by absorption in the EBL.

This paper is organized as follows. Section \ref{datasample} describes the MAGIC and \lat\ datasets, including the data selection and the analysis methods for both instruments. In section \ref{density_constraints} we introduce the proposed  methodology to measure the EBL density assuming different EBL template models, and present the results, both for the full sample and for four subsamples defined by source redshift. This section describes also the systematic uncertainties of the method. Section \ref{sec_wavelength_resolved_ebl} presents a wavelength-resolved estimate of the EBL density. Finally, in section \ref{summaryandconclusions} we summarize the main results of this study, and Appendix A provides technical details of the analysis method used throughout the paper.
\section{Data sample}\label{datasample}
\begin{figure}
	\includegraphics[width=\columnwidth]{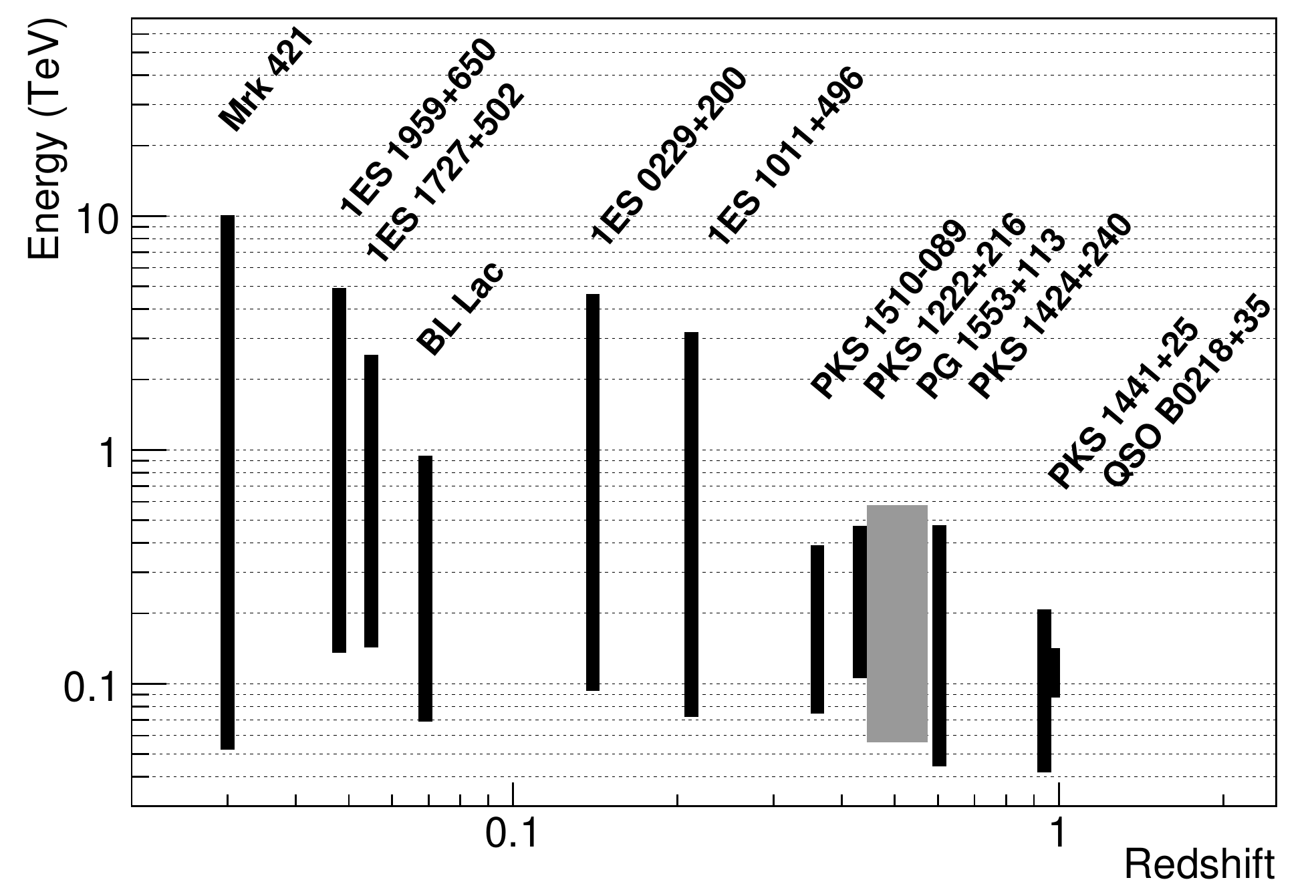}
    \caption{Summary of the MAGIC data sample: energy range probed by the observations vs. source redshift.}
    \label{Erange_vs_z}
\end{figure}

A large majority of the known extragalactic VHE sources are blazars, a class of active galactic nuclei with jets closely aligned with the line of sight of the observer \citep{1995PASP..107..803U}. For the present study we have selected 32 VHE spectra from 12 blazars, obtained with MAGIC in the period June 2010 - May 2016, with a total observation time (after quality cuts) of 316 hours. The sources span the range 0.030 - 0.944 in redshift, and the sample includes both multi-year observations of persistent sources (1ES~0229+200, PG~1553+113, PKS~1424+240) and target of opportunity observations of flaring sources (on timescales from less than one hour to around one month). Table~\ref{tab:EBLsampleTable} lists the twelve sources and provides the basic parameters of the observations. Figure~\ref{Erange_vs_z} shows the energy range of the MAGIC observations for each source, plotted versus the source redshift.
\begin{table}
\caption{Summary of the 32 MAGIC spectra used in the determination of the EBL density. The sample includes flat spectrum radio quasars (FSRQs), and intermediate- and high-frequency-peaked BL Lac objects (IBLs and HBLs respectively) - see e.g. \citet{2011MNRAS.414.2674G}. The redshift of PG 1553+113 is only approximately known; the quoted estimated range is from \citet{2010ApJ...720..976D}} 
\label{tab:EBLsampleTable}
\begin{center}
\begin{tabular}{@{}llclr@{}}
\hline
\multirow{2}{*}{Source}  & Blazar & \multirow{2}{*}{Redshift} & Observational & Obs. \\
& Type & & Period & t (h) \\
\hline
Markarian 421 & \multirow{2}{*}{HBL} & \multirow{2}{*}{0.030} & 2013.04.10-19, & \multirow{2}{*}{43.8} \\
 (15 datasets) & & & 2014.04.26  \\
\noalign{\vskip 1mm}
1ES 1959+650 & HBL & 0.048 & 2015.11.06-18 & 4.8 \\
\noalign{\vskip 1mm}
\multirow{2}{*}{1ES\,1727+502} & \multirow{2}{*}{HBL} & \multirow{2}{*}{0.055} & 2015.10.12- & \multirow{2}{*}{6.4} \\
& & & 2015.11.02 \\
\noalign{\vskip 1mm}
BL Lacertae & IBL & 0.069 & 2015.06.15 & 1.0 \\
\hline
1ES 0229+200 & HBL & 0.14 & 2012-2015 & 105.2 \\
\noalign{\vskip 1mm}
\multirow{2}{*}{1ES 1011+496} & \multirow{2}{*}{HBL} & \multirow{2}{*}{0.212} & 2014.02.06- & \multirow{2}{*}{11.8} \\
& & & 2014.03.07 \\
\hline
PKS 1510$-$089 & \multirow{2}{*}{FSRQ} & \multirow{2}{*}{0.361} & 2015.05.18-19, & \multirow{2}{*}{5.0} \\
(2 datasets) & & & 2016.05.31 \\
\noalign{\vskip 1mm}
PKS 1222+216 & FSRQ & 0.432 & 2010.06.18 & 0.5 \\
\noalign{\vskip 1mm}
PG 1553+113 & \multirow{2}{*}{HBL} & 0.43- & \multirow{2}{*}{2012-2016} & \multirow{2}{*}{66.4} \\
(5 datasets) & & 0.58 \\
\hline
PKS 1424+240 & \multirow{2}{*}{HBL} & \multirow{2}{*}{0.604} & \multirow{2}{*}{2014, 2015} & \multirow{2}{*}{49.1} \\
(2 datasets) \\
\noalign{\vskip 1mm}
PKS 1441+25 & FSRQ & 0.939 & 2015.04.18-23 & 20.1 \\
\noalign{\vskip 1mm}
QSO B0218+35 & FSRQ & 0.944 & 2014.07.25-26 & 2.1 \\
\hline
Total: & & & & 316.1 \\
\end{tabular}
\end{center}
\end{table}

\subsection{MAGIC observations}
MAGIC is a system of two IACTs located at the Roque de los Muchachos Observatory on the island of La Palma in Spain \citep{ALEKSIC201661}. Equipped with 17 m - diameter mirror dishes and fast, 1039-pixel PMT cameras, the telescopes record images of extensive air showers in stereoscopic mode, enabling the observation of VHE gamma-ray sources at energies $\gtrsim 50$ GeV. The data analysis is performed using the standard MAGIC analysis and reconstruction software MARS \citep{zaninMARS,MAGICperformance2016}. All data used for the present study were taken during dark nights in good weather conditions. Atmospheric transmission was monitored with the MAGIC LIDAR \citep{lidar}. After data  quality cuts, the median of the aerosol transparency measurements within each of the 32 samples in Table~\ref{tab:EBLsampleTable}, relative to that of an optimal night, ranges between 0.9 and 1.0, except for the case of the observations of PKS 1510$-$089 on May 2015, for which it is 0.83 (this is the only sample for which a correction for atmospheric transmission, based on the LIDAR data, had to be applied).  
\par
After selection of pixels with a significant signal in each of the cameras, a set of parameters describing the images is calculated (among which are the well-known Hillas parameters, \citealt{1985ICRC....3..445H}). The stereoscopic reconstruction of the geometry of the shower is then performed, using the parameters from both images, to obtain its direction and location relative to the telescopes. The energy of the primary is estimated, assuming a purely electromagnetic shower, using look-up tables which make use of all relevant parameters. This method allows us to achieve a relative energy resolution between 15 and 23\% depending on the energy \citep{MAGICperformance2016}. The random forest method \citep{Breiman2001}, fed with image parameters, is then used to obtain a refined estimate of the shower direction, and to tag events with a test statistic for particle identification dubbed {\it hadronness} \citep{randomforest}. Energy-dependent cuts in {\it hadronness}, and in the angular distance between the target source and the reconstructed event direction, are then applied to improve the signal to noise ratio of the data before obtaining the VHE gamma-ray spectrum of the observed source. In the process outlined above, the energy look-up tables and random forests are created using a training sample of Monte Carlo (MC) simulated gamma-ray initiated showers (using version 6.500 of the CORSIKA~program, \citealt{corsika}, and a detailed simulation of the optics and electronics of the telescopes). For the training of the event-tagging random forest, a sample of hadronic-shower-dominated real MAGIC data from off-source observations is used together with the gamma MC. An independent sample of MC gamma events is processed in the same way as the real data to obtain the instrument response functions (effective area and energy migration matrix) needed for the spectral analysis of the sources. Since the data spans multiple years and the performance of MAGIC has changed over time, several independent MC libraries (tuned to the MAGIC performance in independent periods lasting from few months to over one year) were used in the analysis.

\subsection{\lat\ observations}
\label{latobservations}
The {\em Fermi} Large Area Telescope (LAT, \citealt{atwood09}) is a pair conversion detector consisting of a $4\times4$ array of silicon strip trackers and tungsten converters and a Cesium Iodide (CsI) based calorimeter. The instrument is fully covered by a segmented anti-coincidence shield which provides a highly efficient vetoing against charged particle background events. The LAT is sensitive to gamma rays from $20\,{\rm MeV}$ to more than $300\,{\rm GeV}$. It normally operates in survey mode, covering the whole sky every three hours and providing an instantaneous field of view of $2.4\,{\rm sr}$ (that is, $20\%$ of the sky).

\begin{table*}
    \begin{center}
	\caption[List of observations selected in \lat.]{List of observations selected in \lat. Period stands for the corresponding MAGIC observation. For 1ES\,0229+200, {\em all} means that all the data available from the \lat\ was integrated. For PG\,1553+113, the periods dubbed ST0X0Y group data taken in periods of stable MAGIC performance and IRFs, where X denotes major hardware changes and Y refers to minor changes. They correspond to the 5 datasets introduced in Table~\ref{tab:EBLsampleTable}. TSTART and TSTOP denote the limits of the \lat\ integration periods. Note that for some sources, periods within these limits for which no MAGIC observations exist have been excluded. The quoted {\em Redshift} and Analysis {\em Model} are those used in fitting the \lat\ data of the given source, from which the bow-ties later used in the EBL constraints are obtained. Finally, {\em TS} denotes the Test Statistics, related to the statistical significance of the source detection.}  
    \label{tab:LATexposures}
	{\fontsize{8.5}{10}\selectfont
		\begin{tabular}{@{}lrllrlr@{}}
			\hline
			Source [Period]       &   Redshift & TSTART              & TSTOP               &   Exposure (d) & Model &   TS \\
			\hline
			1ES\,0229+200 [all]   &      0.14  & 2009-11-01T00:00 & 2017-01-01T12:00 &  2200     & PWL   &  113 \\
			1ES\,1011+496 [2014]  &      0.212 & 2014-02-05T12:00 & 2014-03-07T12:00    &  17.7 & EPWL &  426 \\
			1ES\,1727+502 [2015]  &      0.055 & 2015-03-29T12:00 & 2015-11-02T12:00 &    57.3   & PWL  &   98 \\
			1ES\,1959+650 [2015]  &      0.047 & 2015-11-05T12:00 & 2015-11-18T12:00 &    11     & LP   &  405 \\
			B\,0218+357 [2014]    &      0.944 & 2014-07-24T21:00 & 2014-07-26T12:00 &     1.37  & PWL  &  179 \\
			BL\,Lac [20150615]    &      0.069 & 2015-06-14T15:00 & 2015-06-15T03:00 &     0.376 & PWL  &   26 \\
			Mrk\,421 [20130410]   &      0.03  & 2013-04-09T12:00 & 2013-04-10T12:00 &     0.845 & PWL  &  179 \\
			Mrk\,421 [20130411]   &      0.03  & 2013-04-10T18:00 & 2013-04-11T06:00 &     0.389 & PWL  &   44 \\
			Mrk\,421 [20130412]   &      0.03  & 2013-04-11T18:00 & 2013-04-12T06:00 &     0.388 & PWL  &  120 \\
			Mrk\,421 [20130413a]  &      0.03  & 2013-04-12T12:00 & 2013-04-13T12:00 &     0.848 & PWL  &  158 \\
			Mrk\,421 [20130413b]  &      0.03  & 2013-04-12T12:00 & 2013-04-13T12:00 &     0.848 & PWL  &  158 \\
			Mrk\,421 [20130413c]  &      0.03  & 2013-04-12T12:00 & 2013-04-13T12:00 &     0.848 & PWL  &  158 \\
			Mrk\,421 [20130414]   &      0.03  & 2013-04-13T12:00 & 2013-04-14T12:00 &     0.844 & PWL  &  122 \\
			Mrk\,421 [20130415a]  &      0.03  & 2013-04-14T21:17 & 2013-04-15T04:13 &     0.209 & PWL  &   81 \\
			Mrk\,421 [20130415b]  &      0.03  & 2013-04-14T21:17 & 2013-04-15T04:13 &     0.209 & PWL  &   81 \\
			Mrk\,421 [20130415c]  &      0.03  & 2013-04-14T21:17 & 2013-04-15T04:13 &     0.209 & PWL  &   81 \\
			Mrk\,421 [20130416]   &      0.03  & 2013-04-15T12:00 & 2013-04-16T09:00 &     0.723 & PWL  &  110 \\
			Mrk\,421 [20130417]   &      0.03  & 2013-04-16T18:00 & 2013-04-17T06:00 &     0.359 & PWL  &   23 \\
			Mrk\,421 [20130418]   &      0.03  & 2013-04-17T12:00 & 2013-04-18T12:00 &     0.845 & PWL  &   87 \\
			Mrk\,421 [20130419]   &      0.03  & 2013-04-18T12:00 & 2013-04-19T12:00 &     0.844 & PWL  &  104 \\
			Mrk\,421 [2014]       &      0.03  & 2014-04-25T18:00 & 2014-04-26T06:00 &     0.365 & PWL  &   69 \\
			PG\,1553+113 [ST0202] &      0.45  & 2012-02-28T12:00 & 2012-03-04T12:00 &     4.22  & PWL  &   71 \\
			PG\,1553+113 [ST0203] &      0.45  & 2012-03-13T12:00 & 2012-05-02T12:00 &    41.9   & PWL  &  457 \\
			PG\,1553+113 [ST0302] &      0.45  & 2013-04-07T12:00 & 2013-06-12T12:00 &    55.7   & LP   &  475 \\
			PG\,1553+113 [ST0303] &      0.45  & 2014-03-11T12:00 & 2014-03-25T12:00 &    11.8   & PWL  &  207 \\
			PG\,1553+113 [ST0306] &      0.45  & 2015-01-25T12:00 & 2015-08-07T12:00 &   164     & EPWL & 2606 \\
			PKS\,1222+216 [2010]  &      0.432 & 2010-06-17T20:00 & 2010-06-18T00:00 &     0.152 & LP   &  224 \\
			PKS\,1424+240 [2014]  &      0.6   & 2014-03-23T12:00 & 2014-06-18T12:00 &    73.3   & PWL  &  453 \\
			PKS\,1424+240 [2015]  &      0.6   & 2015-01-22T12:00 & 2015-06-13T12:00 &   120     & PWL  &  945 \\
			PKS\,1441+25 [2015]   &      0.94  & 2015-04-17T12:00 & 2015-04-23T12:00 &     5.06  & PWL  &  621 \\
			PKS\,1510$-$089 [2015]  &     0.36 & 2015-05-17T22:48 & 2015-05-19T02:10 & 0.299 & EPWL &  369 \\
			PKS\,1510$-$089 [2016]  &     0.36 & 2016-05-30T12:00 & 2016-05-31T12:00 & 0.843 & EPWL &  204 \\
			\hline
		\end{tabular}
	}\end{center}
\end{table*}

The \lat\ data were extracted from the weekly LAT data files available at the FSSC data center\footnote{\protect\url{https://fermi.gsfc.nasa.gov/ssc/data/access/}}. For each data sample, we consider only Pass-8 source-class photons detected in a {\it region of interest} (ROI) of  $15^\circ$ radius centered on the nominal position of the analyzed source. 
Only events whose estimated energy lies between $100\,{\rm MeV}$ and $500\,{\rm GeV}$ were selected. Following the event selection recommendations from the \lat\ analysis {\it Cicerone}\footnote{\protect\url{https://fermi.gsfc.nasa.gov/ssc/data/analysis/documentation/Cicerone/}}, we only included good data ({\tt (DATA\_QUAL>0)\&\&(LAT\_CONFIG==1)}) with zenith distance lower than $90^\circ$. 
The time-based filtering of the data was done to balance out photon statistics and systematic uncertainties arising from the lack of strict simultaneity with respect to the MAGIC observations. By default, we selected events from 12:00 UTC (noon) of the day preceding the first VHE observations until 12:00 UTC of the day following the last night in which VHE data were taken, ensuring that at least 24 hours of LAT data are included in the analysis. For very fast flares with enough photon statistics in high energy gamma rays, we further restricted the time intervals to ensure that MAGIC and \lat\ data corresponded approximately to the same level of activity of the source. This included $3.6\,{\rm h}$ centered around the two MAGIC observations of PKS~1510$-$089 in 2015, $6\,{\rm h}$ centered around the MAGIC observations of PKS~1222+21, $8\,{\rm h}$ and $7\,{\rm h}$ centered around the MAGIC observations of Mrk~421 on April 11th and April 15th (2013) respectively. For the two highest-flux nights during the April 2013 Mrk421 flare (20130413 and 20130415), the MAGIC observations were split into three sub-samples (a,b,c), each of $\simeq 2$ h duration, according to their flux level. The signal in the LAT observations was however not high enough to provide independent spectra for each of those  sub-periods, and hence a single LAT spectrum has been computed for each night. It must be noted that, while the gamma-ray flux measured by MAGIC on those nights is highly variable (by up to a factor two) in the TeV range, it is stable within uncertainties around 100 GeV, so all three MAGIC spectra of each night connect smoothly with the corresponding average LAT spectrum.

\par
The case of 1ES~0229+200 also demanded a special treatment. The source is an extreme HBL BL Lac which required the integration of a much larger LAT exposure, of more than $6$ years, in order to provide a reasonable detection, ${\rm TS}\sim 80$. ${\rm TS}$ is a Test Statistic for source detection defined in terms of a likelihood ratio test (LRT) as ${\rm TS} = -2\log(L_{\rm max,H0}/L_{\rm max,H1})$, where H0 is the null hypothesis, obtained by removing the source of interest from the source model that was generated for H1 \citep{mattox96}. Finally, for 1ES~1011+496, \lat\ observations were optimized to account for the MAGIC Moon break, hence including only data from February 5th to February 12th (2012) and then February 21st until March 7th. 
\par
\begin{figure*}
	\includegraphics[width=1\columnwidth]{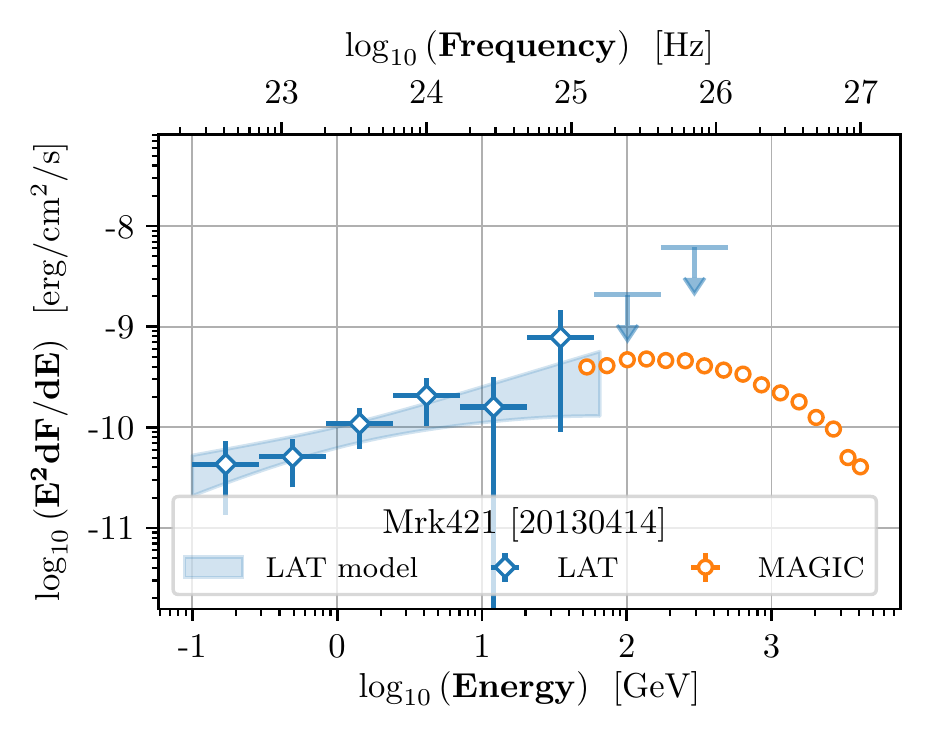}
    \includegraphics[width=1\columnwidth]{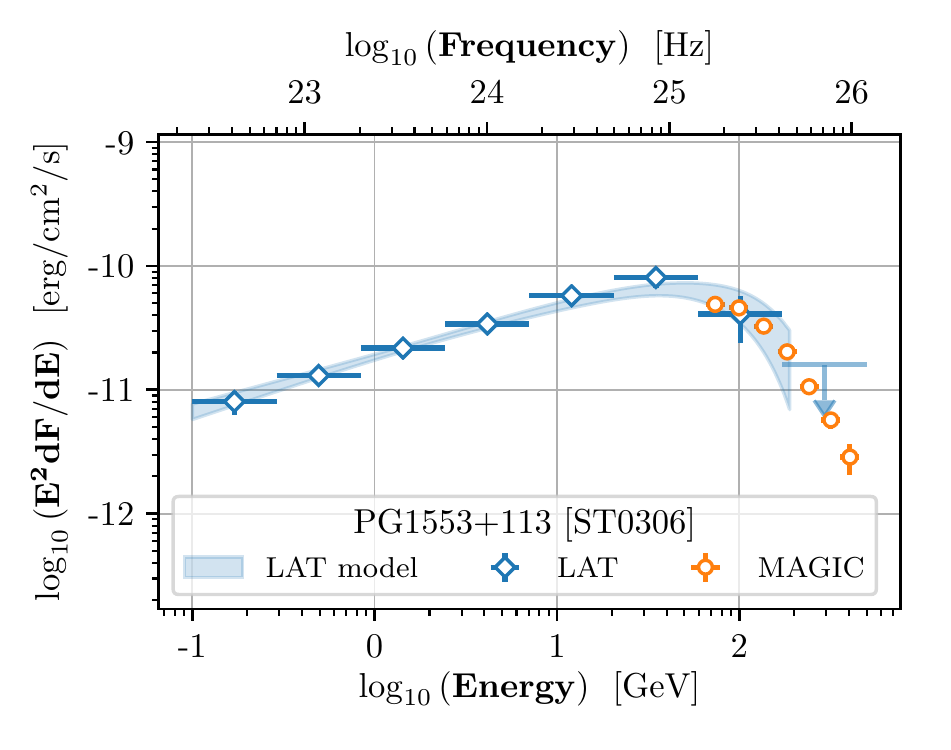}
    \caption{Detailed broadband gamma-ray spectral energy distributions of the Markarian 421 dataset of 2013 April 14 (short exposure) and the ST0306 2015 data from PG~1553+113 (long integration time), showing the good level of agreement achieved for both the MAGIC spectral points (orange open points) and the HE bow-ties and spectral points obtained through the maximum likelihood analysis of \lat\ data (blue). The y-axes correspond to {\it observed} fluxes, i.e., they include the effect of absorption by the EBL. LAT upper limits are for $2\sigma$ confidence level.}
    \label{SEDs_MAGICLATpoints}
\end{figure*}

The full list of \lat\ observations is shown in Table~\ref{tab:LATexposures}. Two examples of SEDs obtained in contemporaneous MAGIC and LAT observations for Mrk~421 and PG~1553+113 are shown in Figure~\ref{SEDs_MAGICLATpoints}.
For each data sample, the data were reduced and analyzed using the open-source software package {\tt enrico} \citep{Enrico2013} as a wrapper for the {\em Fermi} {\tt ScienceTools} (version v10r0p5)\footnote{\protect\url{http://fermi.gsfc.nasa.gov/ssc/data/analysis/scitools}}. We followed a binned likelihood analysis approach split in PSF event types (0, 1, 2 and 3) with 10 bins per energy decade and using the instrument response functions (IRFs) {P8R2\_SOURCE\_V6}. All the 3FGL (third {\em Fermi} Large Area Telescope source catalog, \citealt{3FGL}) sources within the ROI are included in the model, along with Galactic and isotropic models using {\tt gll\_iem\_v06.fits} and {\tt iso\_P8R2\_SOURCE\_V6\_v06.txt} files respectively. 
The spectral model used for each of the sources was selected in order to maximize $L_{\rm max,H1}$. With the exception of 1ES~0229+200, for which a pure power law was the model of choice, the rest of the data samples were modeled using curved spectral shapes, either by allowing EBL absorption to have an effect at the highest energies or by using models with intrinsic curvature terms (log parabola and power law with exponential cutoff). The spectral parameters of all sources with ${\rm TS}>4$ within a radius of $3^\circ$ around the source of interest were left free in the likelihood maximization. The parameters of the rest of the sources are fixed to the published 3FGL values. We also left free the normalization of the diffuse components. Finally, the data were divided in several energy bins to obtain \lat\ spectral points. The results were found to be in good agreement with those of MAGIC in the overlapping energy range (two examples are shown in Fig. \ref{SEDs_MAGICLATpoints}. Note that the spectral shapes used for the \lat\ analysis described above, and reported in the ``model'' column of Table~\ref{tab:LATexposures}, were chosen based on the LAT data alone ($\mathrm{E}_\gamma \lesssim 100\,\mathrm{GeV}$). They should not be confused with the spectral models used later for the joint analysis of \lat\ and MAGIC data over a wider energy range (see \ref{model selection}).

\section{Constraints on the EBL density}
\label{density_constraints}
In order to set constraints on the EBL from the observed gamma-ray spectra, we have adopted a maximum likelihood approach similar to that used in \cite{abramowski13}. 
We fit simultaneously the 32 spectra in our sample, and use the profile likelihood approach to set constraints on one or more free EBL parameters. In the simplest case, a single EBL parameter $\alpha$ is used to scale the optical depth $\tau(E,z)$ from a given template EBL model. 
We calculate $\tau(E,z)$ from the evolving spectral photon densities provided by the EBL model using Eq.~\ref{eq_tau},
\begin{equation} \label{eq_tau}
        \tau(E,z) = c\int_0^{z} \left|\frac{dt}{dz'}\right| dz' \int_{0} ^2\frac{\mu}{2}d\mu \int^\infty_{\varepsilon_\text{th}} \sigma(\varepsilon,E^\prime,\mu)\, n(\varepsilon,z')\,d\varepsilon
\end{equation}
where $\mu = 1-\cos{\theta}$, with $\theta$ the angle of interaction between the gamma ray and the EBL photon, and $E^\prime = E (1+z^\prime)$ and $\varepsilon$ are their respective energies. The term $|dt/dz^\prime|$ incorporates the $\Lambda$CDM cosmological model, $|dt/dz^\prime|^{-1}=H_0\,(1+z^\prime)\,\sqrt{\Omega_m(1+z^\prime)^3+\Omega_\Lambda}$, for which we adopted $H_0=70$~km/s/Mpc, $\Omega_{m}=0.3$, and $\Omega_{\Lambda}=0.7$. The factor $n(\epsilon, z^\prime)$ is the proper number density of EBL photons per unit energy. Finally, $\varepsilon_\text{th}$ is the EBL photon energy threshold for the pair production process, $\varepsilon_\text{th} = 2 m_e^2 c^4 / (\mu \ E^\prime)$, and $\sigma(\varepsilon,E^\prime,\mu)$ is the cross section of the process \citep{Heitler}.
\par
The absorption of VHE photons can be described by a term $e^{-\alpha \tau(E,z)}$, which depends on the energy $E$ of the gamma rays and the redshift $z$ of the source. The spectrum of gamma rays arriving at Earth from the source can then be written as $dF/dE = (dF/dE)_{\rm intrinsic} \; e^{-\alpha \tau(E,z)}$. The spectrum is then folded with the MAGIC IRFs (effective area and energy migration matrix) derived from Monte Carlo simulations, and multiplied by the effective observation time, to obtain the expected number of detected gamma-ray events as a function of the estimated energy. These values and the actually observed numbers of events in bins of estimated energy are then used to build a poissonian likelihood $L$, which is maximized with $\alpha$ as a free parameter. The parameters describing the intrinsic spectra $(dF/dE)_{\rm intrinsic}$ are treated as nuisance parameters. Note that, by scaling $\tau$ by an overall factor $\alpha$ for all $(E, z)$ values, we implicitly assume that both the EBL evolution and spectrum are the ones in the reference model - represented by $n(\varepsilon, z^\prime)$ in eq. \ref{eq_tau}. The formulation of the likelihood $L$ and other technical details of the procedure are explained in appendix \ref{LikelihoodMaxim}. 

\subsection{Maximization of the likelihood
\label{L_maximization}}
The value of the likelihood $L$ is maximized, or rather, $-2 \log L$ minimized, using  the MIGRAD algorithm of ROOT's Minuit2 package \cite{Brun:1997pa,Hatlo:2005cj}. If the maximum achieved likelihood in the space of free parameters is $L_{\rm max}$, in the asymptotic limit, the quantity $-2 \log(L_{\rm max}/L^*)$ is distributed as a $\chi^2$ with the number of degrees of freedom of the problem (the number of fitted $E_{\rm est}$ bins minus the number of free parameters), with $L^*$ being the maximum (unconstrained) possible value of the likelihood, that of a model which predicts exactly the number of recorded ON-source and OFF-source events in every bin of estimated energy. From this $\chi^2$ we can therefore obtain the p-value of the fit.
\par
The profile likelihood of the $\alpha$ parameter, $L(\alpha)$, allows us to obtain the value $\alpha_{\textnormal{best}}$ for which $L$ is maximized, to which we will refer as the ``best-fit'' EBL scale. The  optical depth $\tau$ scales linearly with the EBL density, which means that $\alpha_{\textnormal{best}}$ is also the best-fit EBL density, relative to that of the model. The method can also be interpreted as a LRT between two competing models. In the null hypothesis, the EBL density is fixed to the one in the model, i.e. $\alpha = 1$. The alternative hypothesis has $\alpha$ as an additional free parameter. According to Wilks theorem \citep{wilks1938}, in the asymptotic limit the test statistic $-2 \log \Lambda = -2 \log (L(\alpha=1)/L(\alpha_\textnormal{best}))$ is distributed as a $\chi^2$ with one degree of freedom. This theorem allows us to obtain the $1\,\sigma$ uncertainties in $\alpha_\textnormal{best}$ as the shifts ($\Delta\alpha_+$, $\Delta\alpha_-$) from $\alpha_\textnormal{best}$ which result in $\Delta (-2 \log \Lambda) = 1$.
\subsection{Choice of intrinsic spectral models}
\label{model selection}
An obvious drawback of the method outlined above is the lack of certainty about the intrinsic spectral shapes, $(dF/dE)_{\rm intrinsic}$ of the observed sources. We assume, following authors like \cite{mazin07,abramowski13,biteau15}, that the intrinsic blazar spectra can be described by simple, smooth concave functions with 3 or 4 parameters: power law with exponential or sub/super-exponential cut-off (EPWL, SEPWL), log parabola (LP) and log parabola with exponential cut-off (ELP). Some parameters are limited so that the functions are always concave in the $\log(dF/dE)$ vs. $\log(E)$ representation, i.e., the spectra {\it cannot} become harder for increasing energy. A simple power-law function (PWL, 2 parameters) is also considered as an option, but only for the purpose of estimating the systematic uncertainties (see section \ref{systematics}), since it biases the results towards too high $\alpha$ values (if the intrinsic spectrum is actually concave). The functional expressions for the differential spectra, $dF/dE$, are the following:
\begin{flalign*}
&\text{PWL:} &F_0& \; (E/E_0)^{-\Gamma}\hspace{1cm}
\text{EPWL:}\; F_0 \; (E/E_0)^{-\Gamma}\; e^{-E/E_c}\\
&\text{LP:}  &F_0& \; (E/E_0)^{-\Gamma-b\log (E/E_0)}  \\
&\text{ELP:} &F_0& \; (E/E_0)^{-\Gamma-b\log (E/E_0)}\; e^{-E/E_c}\\
&\text{SEPWL:}\; &F_0& \; (E/E_0)^{-\Gamma}\; e^{-(E/E_c)^d} \\
\end{flalign*}
where $E_0$ is a normalization energy and $F_0$, $\Gamma$, $E_c$, $b$ and $d$ are free parameters.
\par
For a given template EBL model, we scan the values of the scaling factor $\alpha$ between 0 and 2.5 (in steps of 0.05). In each step we try, for each of the spectra, four different intrinsic spectral models: EPWL, LP, ELP and SEPWL. This means that for every spectrum and function we make a likelihood maximization as described in section \ref{L_maximization}, and obtain fit p-values, which allow us to compare how well the different functions describe the data for the given EBL level. We choose the function which provides the best fit (largest p-value) anywhere in the full scanned range of $\alpha$. Alternative selection criteria, such as the one based on the minimization of the Akaike information criterion \citep[AIC,][]{akaike74}, have also been tested and yield similar results. In some cases, LP and EPWL, which have the same number of free parameters, have exactly the same maximum p-value. This occurs when they happen to be degenerate with their common {\it parent function}, a power law. In such cases, we adopt a conservative approach: we choose the function which results in a flatter likelihood curve for $\alpha$, i.e., the one which is most degenerate with the effect of the EBL on the spectrum. In other words, since we assume that either of the two functions is a possible model of the intrinsic spectrum, we choose the one which constrains the EBL {\it less}. It must be noted that other concave functions, not considered by us, could provide an even flatter likelihood and hence a weaker EBL constraint - this underlines the fact that our EBL constraints necessarily rely on the {\it assumption} that the tested spectral models are good enough to describe the intrinsic spectra. 
\par
A first set of intrinsic spectral models is determined following the method just described, and are used to obtain a {\it preliminary} maximum likelihood estimator of the EBL scale $\alpha_{\textnormal{best},0} (+\Delta \alpha_+, -\Delta\alpha_-)$ using all 32 spectra in the sample. A revision of the spectral model selection is then performed following a self-consistent approach, in case of spectra, if any, for which the maximum p-value was found for an $\alpha$ value outside the the range $(\alpha_{\textnormal{best},0} - 2 \Delta \alpha_-, \alpha_{\textnormal{best},0} + 2 \Delta \alpha_+)$. In such cases, the function selection is re-done, this time comparing the p-values in the restricted 2-sigma range around $\alpha_{\textnormal{best},0}$. Then the profile likelihood of $\alpha$ is recalculated with the revised set of spectral models, and the final estimate of $\alpha_\textnormal{best}$ is obtained. This model revision procedure improves (by construction) the p-value of the global fit.
\subsection {Results}
\begin{figure}
	\includegraphics[width=\columnwidth]{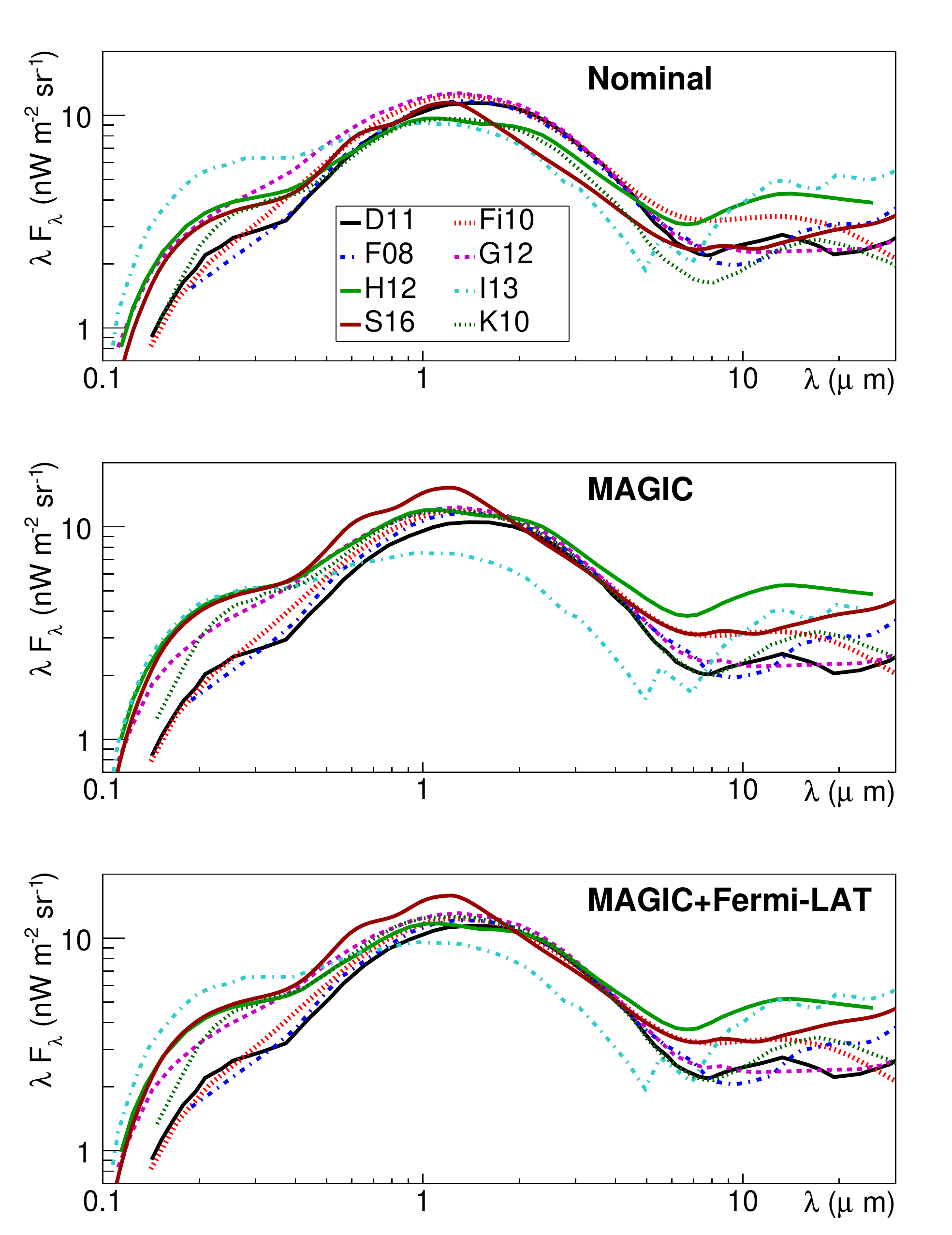}
    \caption{Top panel: SEDs of EBL template models used in this work, see Table \ref{tab:scalefactors}. Middle and bottom panels: the same EBL SEDs scaled by the best-fit EBL scale factors obtained through the analysis of MAGIC-only and MAGIC+\lat\ data respectively.}
    \label{ScaledModelsSED}
\end{figure}

The method has been applied both to the MAGIC data alone, and to a combination of the MAGIC data and the \lat\ data (the latter in the form of spectral {\it bow-ties}, i.e. flux and photon index at a given energy, with their respective uncertainties) which help constrain the intrinsic spectral parameters of the sources, as explained in appendix \ref{LikelihoodWithFermi}. The analysis was repeated for eight different template EBL models - see Table~\ref{tab:scalefactors} for the references and correspondence to the short names we will use to refer to them. 
The spectral energy distribution of the EBL (at $z=0$) between 0.1 and 30 ${\rm \mu m}$ according to the eight models is shown on the top panel of Figure~\ref{ScaledModelsSED}. We think that these eight models are a good representation of the state-of-the-art in EBL research. They span the whole range of four categories (or methodologies) described in the Introduction, i.e. (1) forward evolution, (2) backward evolution, (3) inferred evolution, (4) observed evolution.
\par
\begin{figure}
	\includegraphics[width=\columnwidth]{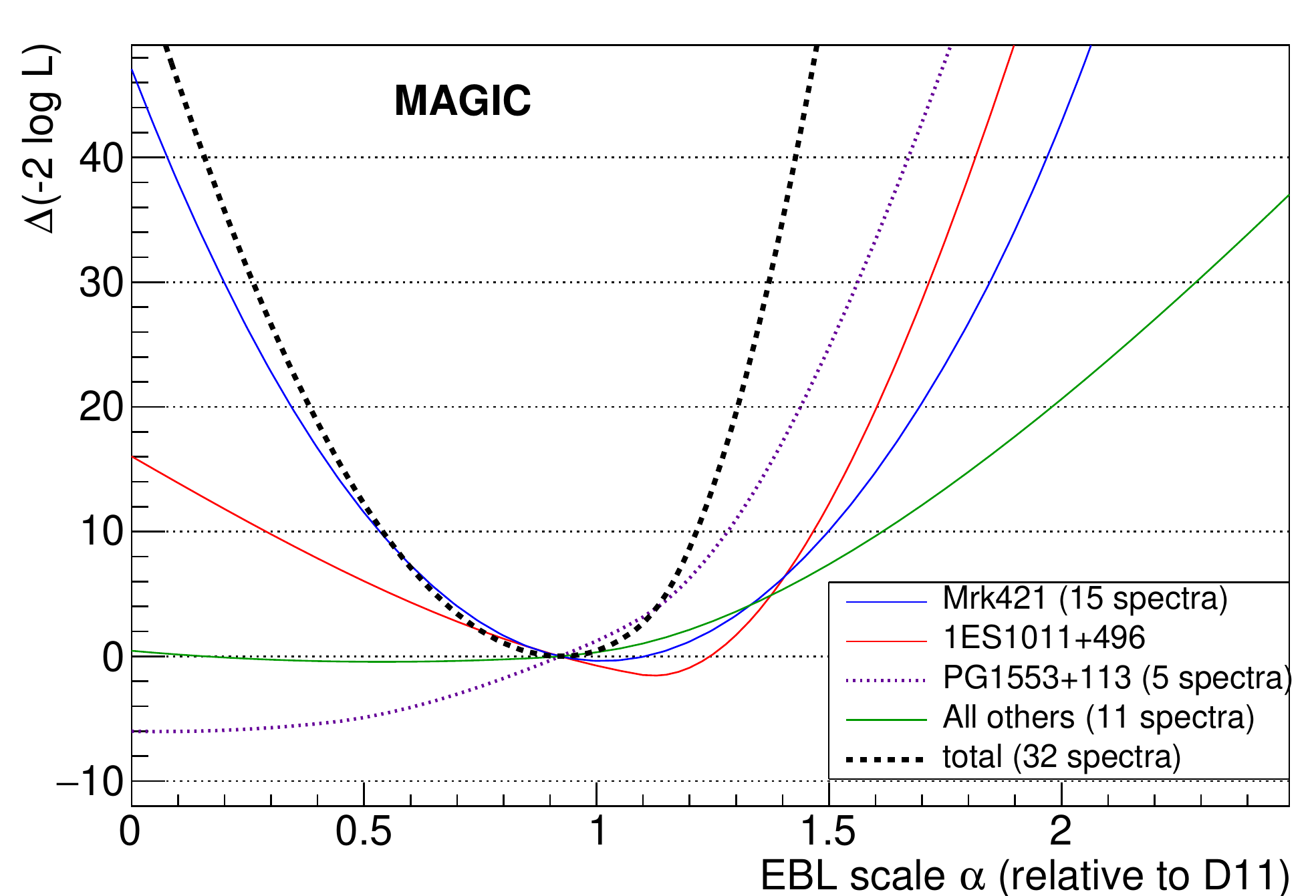}
	\includegraphics[width=\columnwidth]{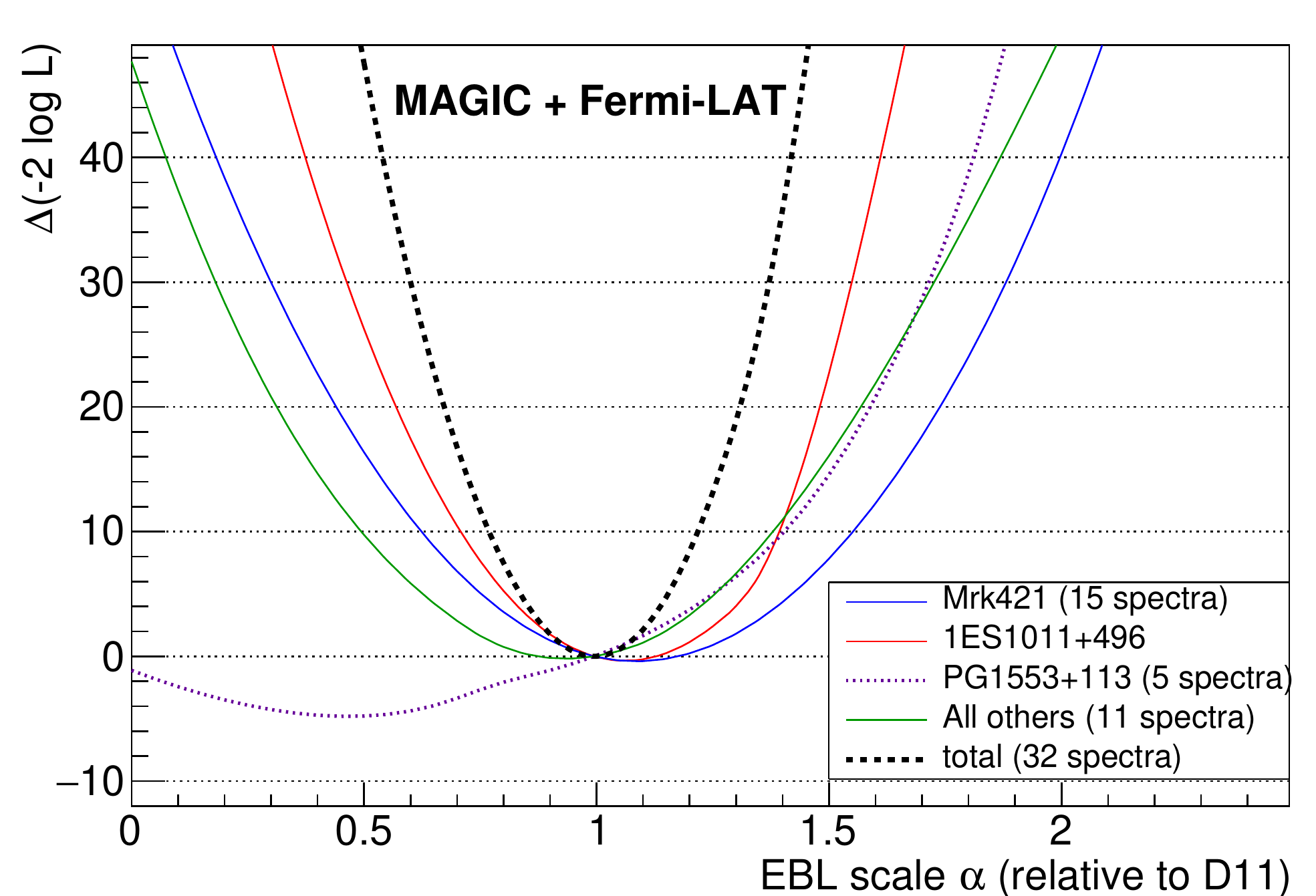}
    \caption{Profile likelihood of the EBL scale relative to the D11 template, for the joint analysis of 32 spectra (dashed black curves) using MAGIC-only and MAGIC+\lat\ data. The colored curves are the profile likelihoods obtained with subsets of the 32 spectra.}
    \label{ProfileLikelihood}
\end{figure}
\begin{figure*}
	\includegraphics[width=2\columnwidth]{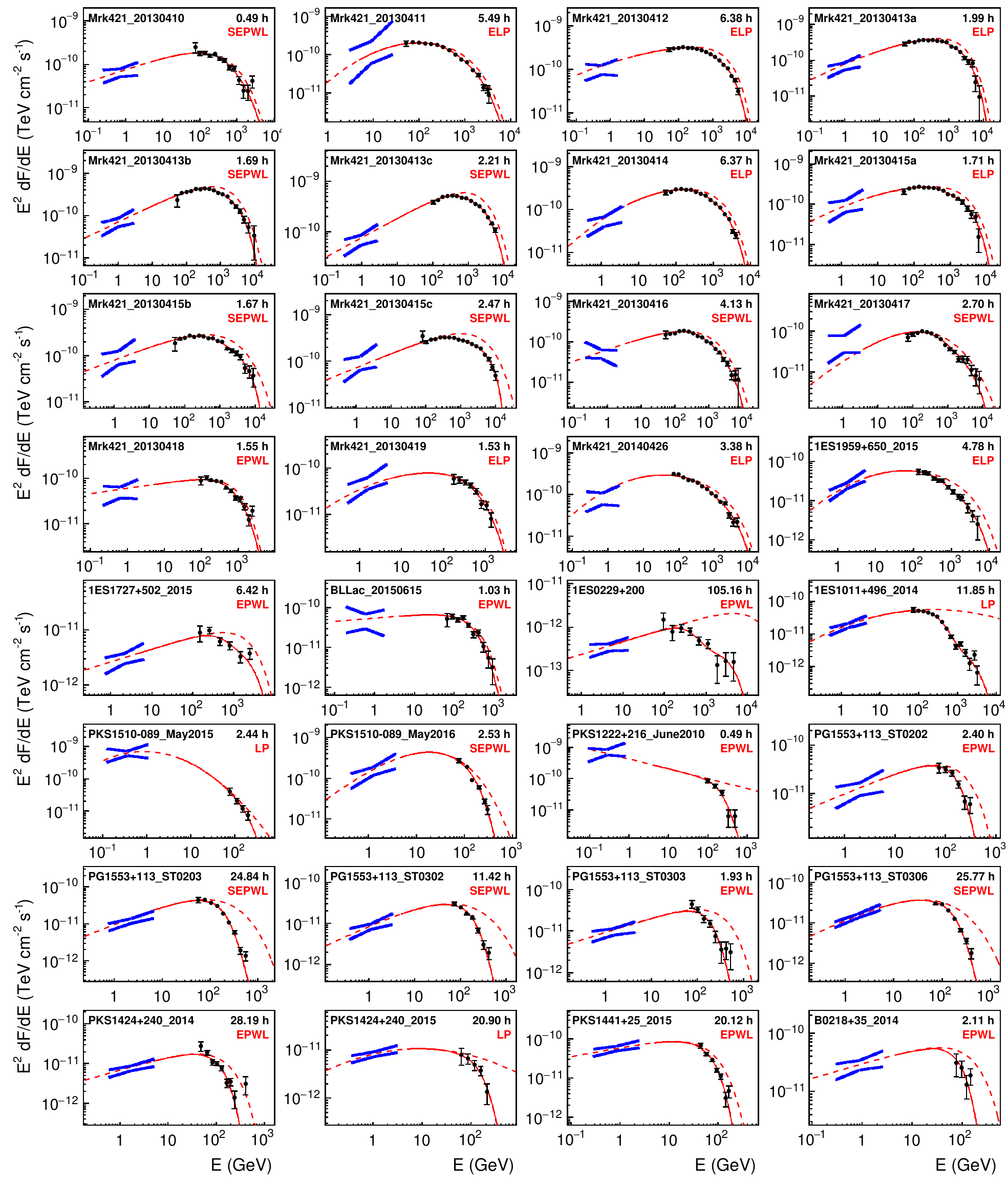}
    \caption{Spectral energy distributions of the 32 spectra measured by MAGIC (black points) and \lat\ (blue bow-ties). The fits correspond to the analysis which uses the D11 EBL template and a single free EBL parameter (overall scale factor) - see Table~\ref{tab:scalefactors}. The dashed red curves are the best-fit intrinsic spectra, the functional form of which is shown in red (as an acronym) below the observation time. The solid curves are the corresponding absorbed spectra. Each of the individual spectral points (black dots) is obtained from the excess of gamma-like events in a given bin of {\it estimated} energy $E_{\rm est}$; the corresponding flux is evaluated at the median {\it true} energy of the events (as estimated from the Monte Carlo simulation).}
    \label{SEDs_D11}
\end{figure*}

Figure \ref{ProfileLikelihood} shows the profile likelihood curves from which the best-fit EBL scale factors have been obtained for the case of the \cite{dominguez11a} model (hereafter, D11). Table~\ref{tab:scalefactors} presents the best-fit scale factors for each of the EBL templates, and the associated statistical uncertainties. Note that, as expected, the uncertainties are smaller when the \lat\ data are included in the analysis. This improvement comes together with a stronger assumption on the intrinsic source spectra, namely that they are well represented by the simple concave functions listed in section \ref{model selection} over a wider energy range, spanning both the \lat\ and MAGIC bands. In particular, this has an effect on the estimated p-values, which are around 20 times smaller when \lat\ data is included. It should be remarked that the small p-values in Table~\ref{tab:scalefactors} (all of them smaller than 0.02) are not surprising, given that (i) the method assumes no uncertainties in the energy- and redshift-dependence of the optical depths from the template EBL model; (ii) the true underlying spectra of the sources may be more complex that the used models; and (iii) no instrumental systematic uncertainties are yet considered - they are treated separately in section \ref{systematics}. Despite these caveats, a reasonable agreement between the MAGIC and LAT spectra is achieved for all the analyzed samples, as can be seen in Fig. \ref{SEDs_D11}.
\par
For half of the tested EBL models (D11, Fi10, F08 and G12) the best-fit scale factors are compatible with 1.0 within $1\,\sigma$, meaning that our data are compatible with the EBL density in the models. Even considering only the statistical uncertainties, the data do not allow to discriminate among the four EBL models. The other four templates (H12, I13, S16 and K10) result in scale factors which are between 1.4 and $2.2\,\sigma$ away from $\alpha=1.0$ for the MAGIC-only analysis (and between 2.3 and $3.5\,\sigma$ for the analysis including \lat, with the only exception of I13, in which the best-fit is compatible with the EBL density in the model - though it also has the worst p-value of the whole set). It must be remarked that the K10 model accounts only for the contribution to the EBL of resolved sources, and is presented by their authors as a ``minimal'' EBL model in the optical and near infrared bands, the spectral range to which our gamma-ray data is most sensitive. It is therefore natural that for such a model a best-fit scale larger than one is obtained. Likewise, the H12 model is based on measurements only up to 24 $\mu m$, so we can expect it to underestimate the total optical depth for the highest energy observations in our sample - which in turn results in a best-fit scale factor larger than 1.
\begin{table*}
\begin{center}
\caption{EBL density constraints (best-fit EBL scale factor $\alpha_{\rm best}$) using MAGIC and MAGIC + \lat\ spectra.\label{tab:scalefactors}}
\begin{tabular}{lllllll}
\multicolumn{1}{c|}{} &   \multicolumn{3}{|c||}{------ MAGIC-only analysis ------} & \multicolumn{3}{|c||}{------ MAGIC + \lat\ analysis ------} \\
\hline
\multirow{2}{*}{EBL template} & Best-fit scale $\alpha_{\rm best}$ & \multirow{2}{*}{$\chi^2$/ndf} & \multirow{2}{*}{p-value} & Best-fit scale & \multirow{2}{*}{$\chi^2$/ndf} & \multirow{2}{*}{p-value} \\
& \multicolumn{1}{|c|}{(stat-only)} & & & \multicolumn{1}{|c|}{(stat-only)} \\
\hline
D11 \cite{dominguez11a} & 0.92 $(+0.11, -0.12)$  & 481/415 & $1.37\times 10^{-2}$  &   1.00 $(+0.07, -0.07)$ & 575/469 & $5.88\times 10^{-4}$ \\
\hline
Fi10 \cite{finke10} & 0.96 $(+0.10, -0.12)$ & 488/416 & $0.83\times 10^{-2}$ &  1.00 $(+0.07, -0.08)$ & 581/472 & $4.43\times 10^{-4}$ \\
\hline
F08 \cite{franceschini08}    & 0.99 $(+0.11, -0.12)$ & 480/415 & $1.50\times 10^{-2}$   &  1.04 $(+0.08, -0.08)$ & 573/469 & $7.34\times 10^{-4}$ \\
\hline
G12 \cite{gilmore12} (fiducial)  & 0.97 $(+0.11, -0.12 )$ & 479/414 & $1.49\times 10^{-2}$  &  1.03 $(+0.08, -0.08)$ & 568/471 & $1.36\times 10^{-3}$ \\
\hline
H12 \cite{helgason12} & 1.24 $(+0.11, -0.16 )$  & 492/417& $0.68\times 10^{-2}$ &  1.21 $(+0.09, -0.10)$ & 582/470 & $3.12\times 10^{-4}$ \\
\hline
I13 \cite{inoue13} & 0.82 $(+0.13, -0.13 )$ & 486/414 & $0.81\times 10^{-2}$ & 1.04 $(+0.11, -0.10 )$ & 595/468 & $0.61\times 10^{-4}$ \\
\hline
S16 \cite{stecker16} & 1.33 $(+0.15, -0.16 )$ & 479/414 & $1.47\times 10^{-2}$ & 1.38 $(+0.11, -0.10 )$ & 569/472 & $1.46\times 10^{-3}$ \\
\hline
K10 \cite{kneiske10} & \multirow{2}{*}{1.23 $(+0.14, -0.15)$} & \multirow{2}{*}{478/415} &  \multirow{2}{*}{$1.69\times 10^{-2}$} & \multirow{2}{*}{1.31 $(+0.09, -0.11)$} & \multirow{2}{*}{566/471} & \multirow{2}{*}{$1.76\times 10^{-3}$} \\
(minimum EBL) \\
\hline
\end{tabular}
\end{center}
\end{table*}
\subsection{Systematic uncertainties}
\label{systematics}
In the results presented above, the only systematic uncertainty that has been considered is $10\%$ in the \lat\ best-fit flux normalization (see appendix \ref{LikelihoodWithFermi}), resulting from the systematic uncertainty in the LAT collection area. This is added in quadrature to its statistical uncertainty, and therefore contributes to the statistical uncertainties in the EBL parameters reported in Table~\ref{tab:scalefactors}. In the rest of this section we discuss systematic uncertainties in the MAGIC results, and how they affect the EBL estimation.
\par
As mentioned above, the EBL estimation method adopted here relies on the assumption that the chosen spectral models are a good representation of the intrinsic gamma-ray spectra of the blazars in the dataset. The derived best-fit EBL density and its statistical uncertainty range are correct only as long as this assumption holds. This is one of the main sources of systematic uncertainty of this method. In order to estimate its effect in our results, we have performed the following tests:
\begin{enumerate}[(i)]
\item Include the power law into the pool of eligible functions.
\item Perform the model selection at a fixed, low-level of EBL density.
\end{enumerate}
When the power law is added to the pool of eligible functions in the process described in section \ref{model selection}, it is preferred to all the others (i.e. yields the highest fit P-value) for some of the spectra. This is the case for between 10 and 15 of the 32 spectra (depending on the EBL template) if only MAGIC data is considered. For the MAGIC and \lat\ analysis, the number drops to between 5 and 8. For those spectra, the additional parameters in the more complex functions do not improve the fit $\chi^2$ enough to compensate for the decrease in the number of degrees of freedom, and so the power law provides the largest fit p-value. Choosing a power law as intrinsic spectral shape has the disadvantage that all the curvature of the {\it observed} spectrum will have to be explained by the EBL, even if part of the curvature is actually intrinsic. This will bias the best-fit EBL scale towards larger values, since the effect will likely go in the same direction for all sources (intrinsic VHE spectra are expected to become generally softer with energy). For this reason we excluded the power law for the EBL estimates reported in Table~\ref{tab:scalefactors}, and we only perform the test (i) in order to estimate the high end of the systematic uncertainty related to the choice of spectral model.
\begin{table}
\begin{center}
\caption{EBL density constraints using MAGIC and MAGIC + \lat\ spectra, including systematic uncertainties.\label{tab:systematicstable}}
\begin{tabular}{lll}
EBL & MAGIC-only & MAGIC + \lat \\
template & (stat+sys) & (stat+sys) \\
\hline
D11  & 0.92 $(+0.23, -0.18)$ & 1.00 $(+0.10, -0.18)$ \\
\hline
Fi10 & 0.96 $(+0.17, -0.28)$ & 1.00 $(+0.09, -0.18)$ \\
\hline
F08  & 0.99 $(+0.21, -0.23)$ & 1.04 $(+0.10, -0.20)$ \\
\hline
G12 & 0.97 $(+0.26, -0.22 )$ & 1.03 $(+0.10, -0.20)$ \\
\hline
H12 & 1.24 $(+0.21, -0.41 )$ & 1.21 $(+0.19, -0.15)$ \\
\hline
I13 & 0.82 $(+0.50, -0.21 )$ & 1.04 $(+0.58, -0.34 )$ \\
\hline
S16 & 1.33 $(+0.34, -0.40 )$ & 1.38 $(+0.28, -0.34 )$ \\
\hline
K10 & 1.23 $(+0.33, -0.30 )$ & 1.31 $(+0.27, -0.23 )$ \\
\hline
\end{tabular}
\end{center}
\end{table}
\par
In a second test (ii) we re-evaluate the model selection (again based on p-values), but fixing the EBL density (for the EBL template being used) at a level determined, at $\lambda = 1.1\; \mu m$, by the galaxy counts measurement in \cite{madau00}: specifically, we use the best-fit value minus  $1\,\sigma$, i.e., $7.81\,{\rm nW\,m^{-2} sr^{-1}}$ (this measurement is shown later in Figure~\ref{WaveDep_D11_vs_direct}). 
By forcing a low EBL density (instead of scanning a wide range) during model selection, we naturally favor more complex functions that can account for part of the observed spectral curvature. The total number of free intrinsic spectral parameters is hence slightly larger, between 2 and 6 more parameters for a total of $\simeq 105$ parameters ($\simeq 115$ for the Fermi+MAGIC analysis), depending on the EBL template. The EBL density estimation with this new set of spectral models will then result in weaker constraints (larger uncertainties) on the low end, due to the larger degeneracy between intrinsic spectra and the effect of the EBL.
\par
\begin{figure}
	\includegraphics[width=\columnwidth]{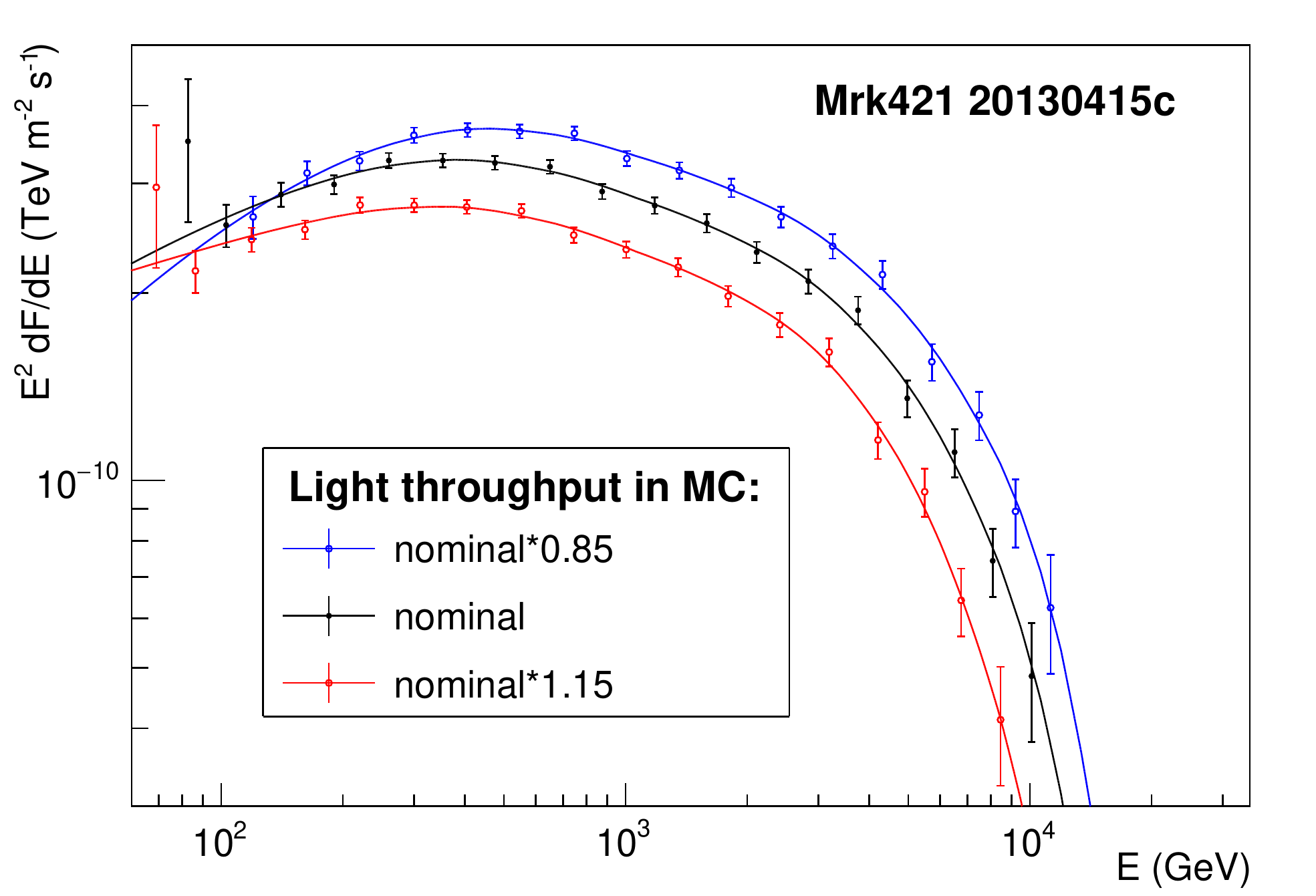}
    \caption{Observed SED of one of the Mrk 421 samples analyzed with three different assumptions on the total overall light throughput of the atmosphere and the telescopes.}
    \label{LightScaleSystematics}
\end{figure}

The other main source of systematic uncertainties we consider, which is related to the IACT observation technique, is the systematic uncertainty in the absolute "energy scale" of the MAGIC telescopes - or, to be more precise, in the total light throughput of the atmosphere and the telescopes. The reconstruction of the energy of gamma rays detected by IACTs fully relies on MC simulations of the shower development in the atmosphere and of the light detection by the telescopes. Any mismatch between the MC-simulated and the actual values of, for instance, the transparency of the atmosphere, or the light collection efficiency of the telescopes, will result in a systematic error in the estimated energy. The MC model is tuned to the characteristics of the telescopes during periods of stable performance (typically lasting several months), and for typical good observation conditions. It is however not tuned to the conditions of each observation night, therefore variations of atmospheric transparency or telescope efficiency {\it within a period} contribute to the {\it statistical} uncertainties reported in the previous section. In order to estimate the effect on the EBL uncertainty of the possible {\it average} data-MC mismatch, we adopt the estimate in \cite{MAGICperformance2016} of a maximum $\pm 15\%$ departure in the absolute energy scale, and 
\begin{enumerate}[(i)]
\setcounter{enumi}{2}
\item re-analyze the whole dataset using spectra reconstructed with MAGIC IRFs corresponding to a total light throughput between 85\% and 115\% of the nominal one, in steps of 5\%  (i.e., 6 different assumptions, besides the case of nominal efficiency).
\end{enumerate}
An example of the effect of those modifications of the IRFs on one of the spectra of the sample is shown on Figure~\ref{LightScaleSystematics}.
\par
The whole EBL estimation procedure, including spectral shape selection, was repeated independently for each of these 6 assumptions on the average MC-data mismatch in light throughput.  The envelope of the $1\,\sigma_{\rm stat}$ statistical uncertainty ranges of the nine different analyses (the default one, the two with modified spectral model selection, and the six with modified light throughput) is taken as the total uncertainty, including systematic uncertainties, reported in Table~\ref{tab:systematicstable}. The total uncertainties are around twice as large (or larger) as statistical uncertainties, showing that this EBL determination method, applied to our data sample, is limited by systematic uncertainties. The lower end of the systematic uncertainty is set, practically in all cases, by the test (ii) described above, hence linked to spectral model selection - the only exception is the I13 template (which is an outlier in terms of EBL spectral shape), for which it is set by the scan of light throughputs (iii). For the upper end of the systematic uncertainty, in contrast, there is no clear pattern: it is sometimes determined by the changes in light throughput (iii), and in other cases by the inclusion of the power-law as as an allowed intrinsic spectral model (i).
\par
A possible additional source of systematic errors is the lack of {\it strict} simultaneity of the MAGIC and \lat\ observations (see section \ref{latobservations}), since source variability may lead to the average emission state being different for the two datasets.  Given the stochastic nature of the behaviour of blazars, however, the systematic errors induced by this mismatch in each of the analyzed spectra will likely affect the EBL estimation in different directions, rather than consistently under- or over-estimate it. This will in turn result in a flattening of the minima of the profile likelihood curves, and an increase of the statistical uncertainties (relative to the ones we would obtain for truly simultaneous MAGIC and \lat\ observations). Beyond that, we currently have no way of estimating the contribution of this effect to the final systematic uncertainty of our measurement.
\subsection{Constraints in bins of redshift}

\begin{figure*}
	\includegraphics[width=1.04\columnwidth]{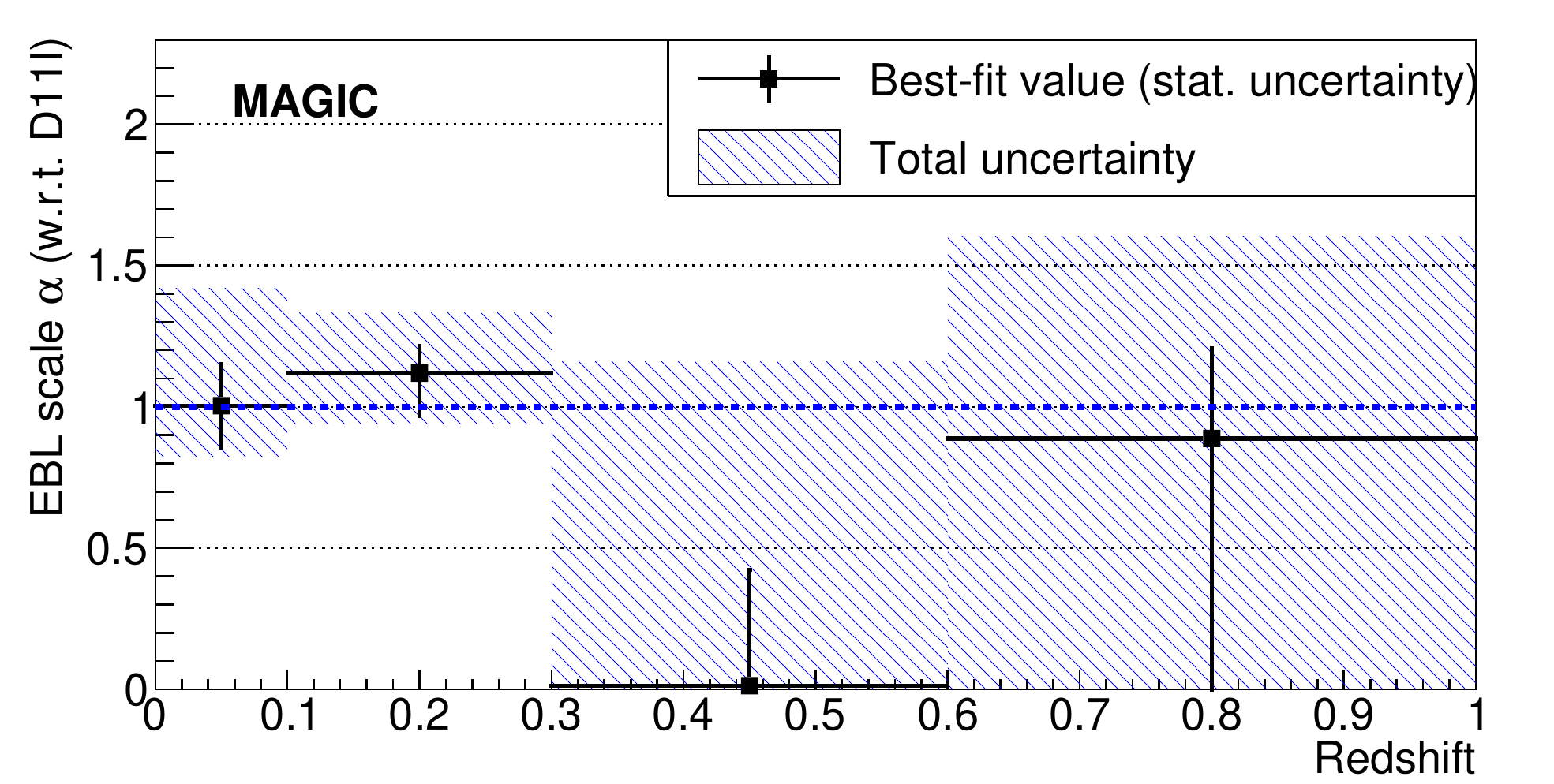}
	\includegraphics[width=1.04\columnwidth]{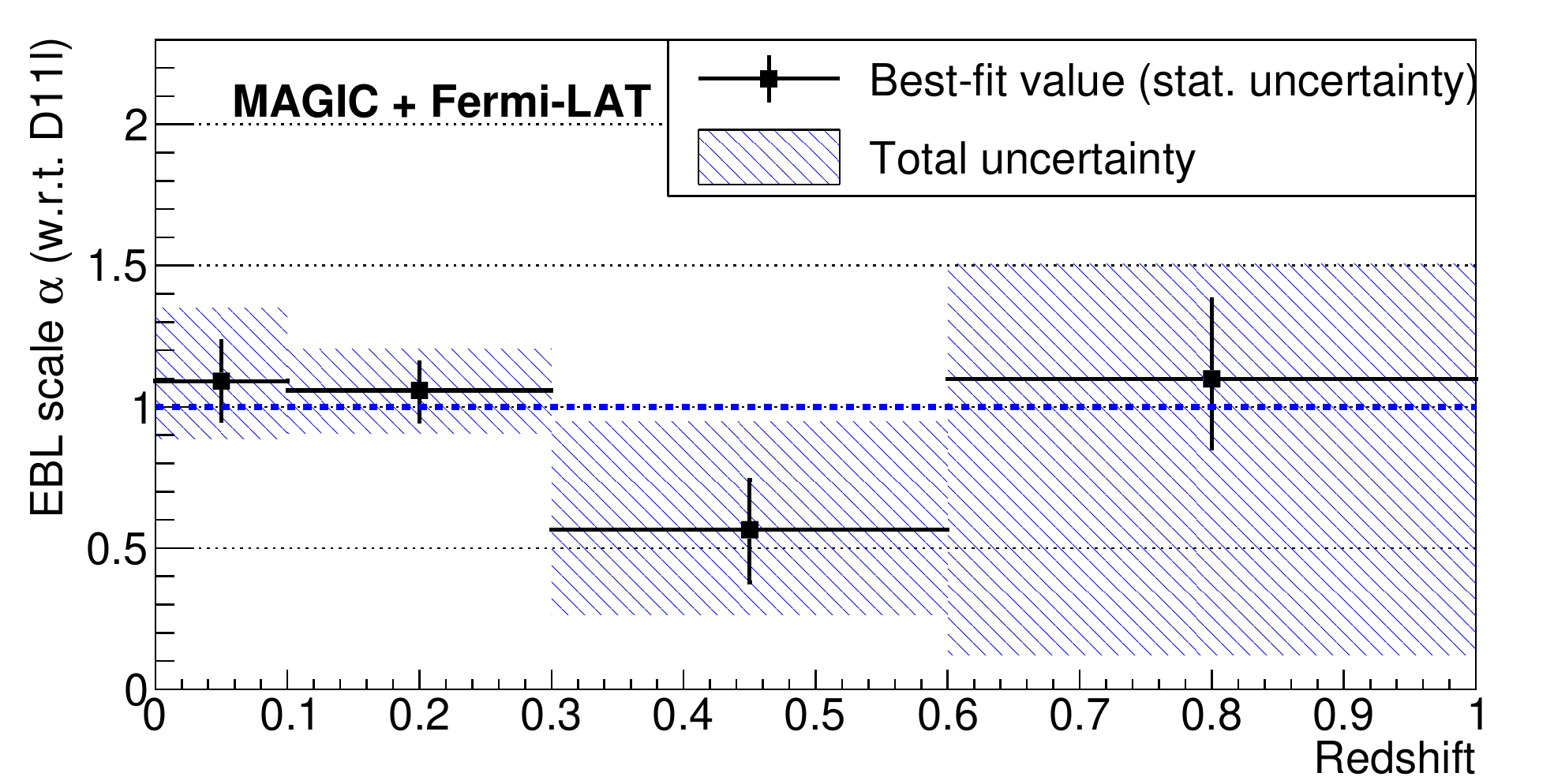}
    \caption{EBL scale, relative to the D11 model, in four bins of redshift. Left panel: MAGIC-only analysis; right panel: MAGIC+Fermi-LAT analysis. The dashed blue band shows the total uncertainty including systematics.}
    \label{Scale_vs_z_D11}
\end{figure*}

\begin{figure}
	\includegraphics[width=1.04\columnwidth]{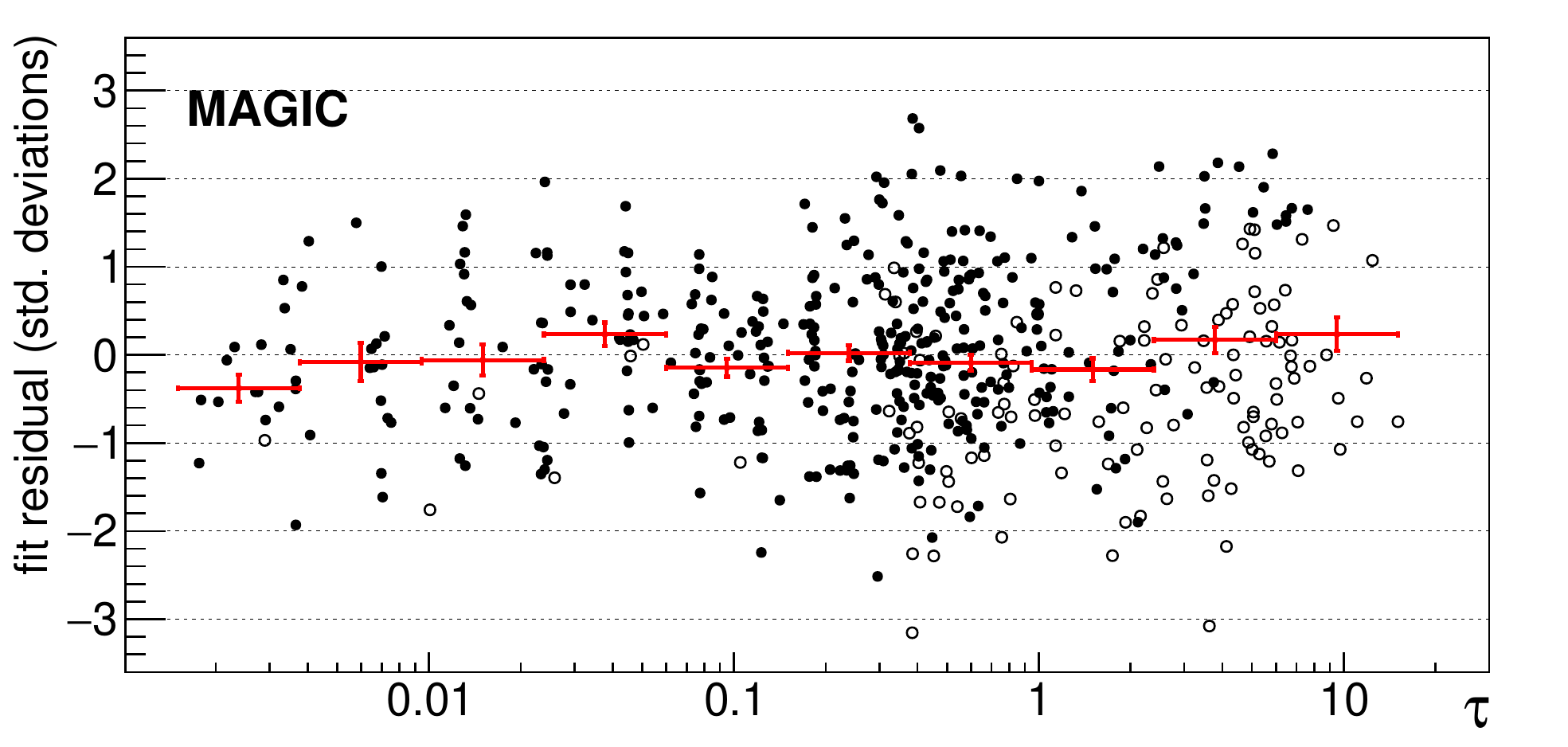}
	\includegraphics[width=1.04\columnwidth]{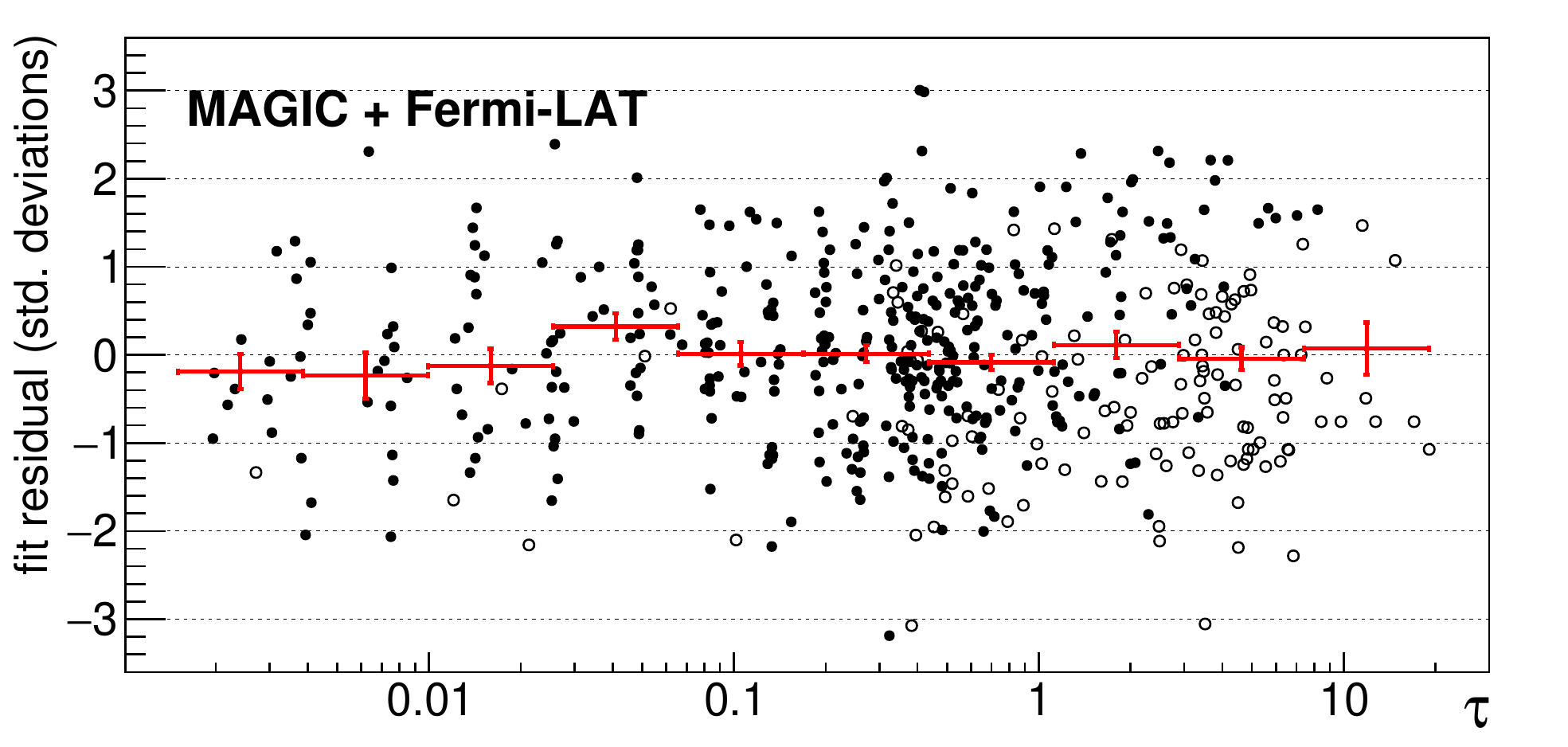}
    \caption{Fit residuals vs. optical depth $\tau$, for the MAGIC data points (i.e. bins of estimated energy). Top panel: MAGIC-only analysis. Bottom panel: MAGIC+\lat\ analysis. Filled symbols indicate bins in which there is a $> 1.5\,\sigma$ excess of gamma-like events above the background fluctuations. The optical depth is calculated for the best-fit EBL scale, relative to the D11 template. The red graph is the average residual in ten bins of $\tau$.
    \label{Residuals}}
\end{figure}
\begin{figure}
	\includegraphics[width=1.04\columnwidth]{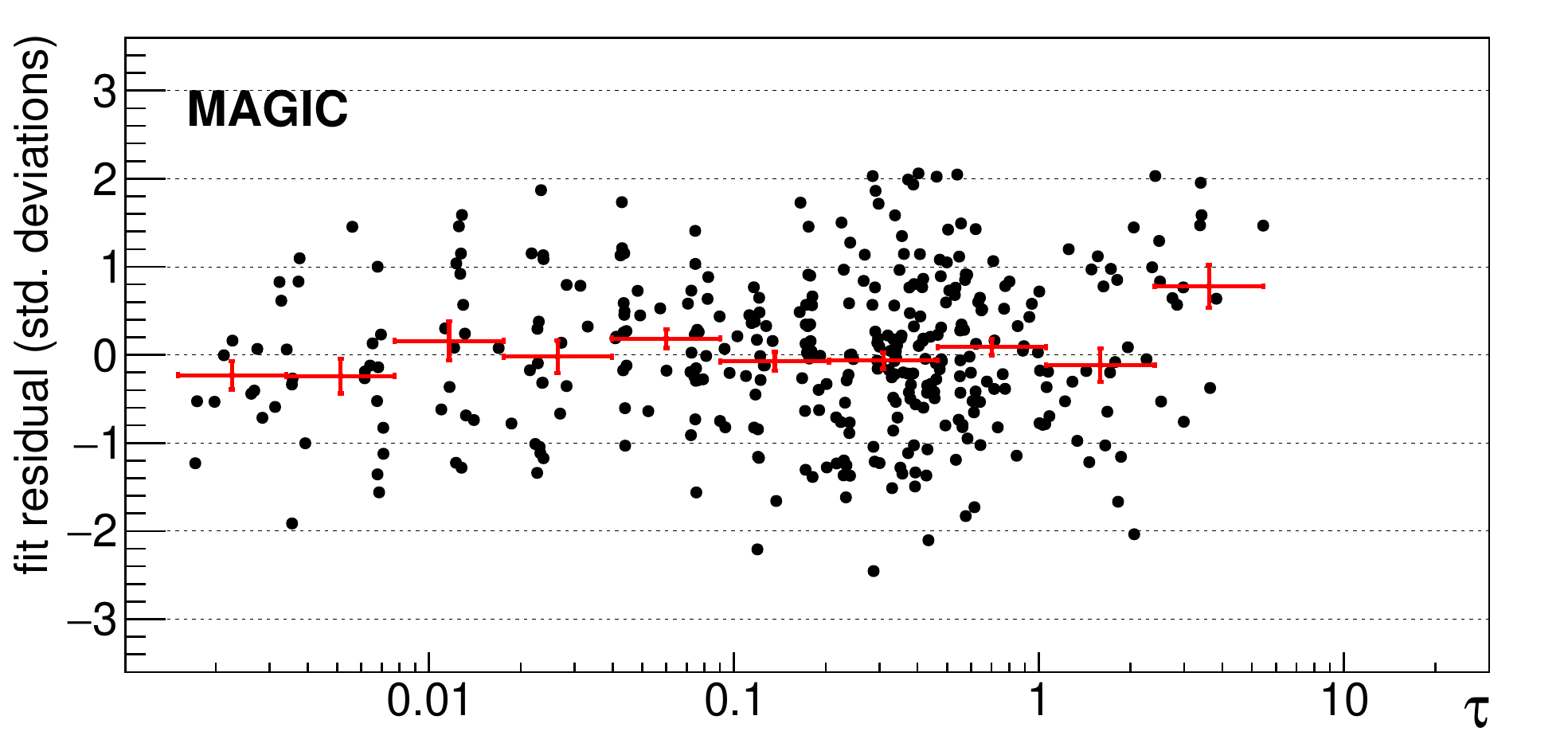}
    \caption{Fit residuals vs. optical depth $\tau$, for the MAGIC-only analysis performed after removing low significance points ($< 1.5 \sigma$, see text).
    \label{Residuals_significancecut}}
\end{figure}
Recent measurements of the star formation history (SFH) are consistent with a strong peak in the star formation rate around $z\sim 2$, decreasing gradually by about one order of magnitude towards $z=0$ as shown by \cite{madau14}. Since the EBL is a tracer of the SFH, any bias in how star formation rate and galaxy evolution are treated in the EBL models could potentially have an effect in our constraints. While the ideal instrument to test the imprint of SFR evolution on gamma-ray blazar spectra is \lat, as it can detect sources up to larger distances, the samples presented in this work at $z\gtrsim 0.5$ are also good candidates to test whether there is any departure in the measured optical depth with respect to the EBL model predictions.
\par
In order to probe the evolution of the EBL, the data were divided in four redshift bins ($0.0-0.1$, $0.1-0.3$, $0.3-0.6$ and $0.6-1.0$, see Table~\ref{tab:EBLsampleTable}) and individual $\alpha_i$ optical depth scaling factors were derived for each bin. The intrinsic spectral models for each of the spectra are the same as in the global EBL scale determination using all redshift bins together (but of course the likelihood maximization is re-done in each bin separately, hence the best-fit spectral parameters are in general different).
For the D11 template, the results are presented in Figure~\ref{Scale_vs_z_D11}. As expected, the strongest constraints are obtained for the two lowest redshift bins, dominated by the high quality spectra of Mrk~421 and 1ES~1011+496 respectively. 
The $3^{\rm rd}$ and $4^{\rm th}$ bins mostly reflect contributions from PG~1553+113 (strong upper bounds), PKS~1424+240 and PKS~1441+25. Only the $3^{\rm rd}$ bin, $0.3<z<0.6$, shows $>1\, \sigma_{\rm stat}$ deviations from $\alpha = 1$ (both for the MAGIC and for the MAGIC+\lat\ analyses). For that redshift range, as well as for $0.6 < z < 1.0$, the effect of EBL attenuation and the intrinsic spectral curvature are hard to disentangle, and the corresponding parameters are degenerate. As a consequence, the expected EBL imprint can be well reproduced with an exponential or super-exponential cut-off, and hence the best-fit $\alpha$, especially for the MAGIC-only analysis, can be well below 1, (even at 0, see left panel of Figure~\ref{Scale_vs_z_D11}). The same effect is visible on the PG 1553 curves in Figure~\ref{ProfileLikelihood}. Nevertheless, the results in all redshift bins are compatible (at the $\simeq 1\,\sigma$ level) with the EBL density in the D11 model, once systematic uncertainties are taken into account. It must be noted that the \lat+MAGIC result in the highest redshift bin, when only statistical uncertainties are considered, represents the first detection of the imprint of the EBL using IACT observations of $z>0.6$ blazars.

\subsection{Fit residuals}

Many extensions of the Standard Model of particle physics, particularly those linked to superstring theories, suggest the existence of light zero-spin bosons commonly known as axion-like particles (ALPs). In the presence of magnetic fields (which exist not only in galaxies, but also on larger scales in the intergalactic space), photon-ALP oscillations are expected to occur if these bosons exist (see e.g. \citealt{2007PhRvD..76l1301D,PhysRevD.76.023001, PhysRevLett.99.231102, 2009PhRvD..79l3511S, 2011PhRvD..84j5030D}). ALPs travel unaffected by interactions with EBL photons, and can oscillate back into VHE photons close to us. This can potentially lead to significant modifications of the effective optical depth $\tau$ that we measure from Earth, modifying the observed source spectra in non-trivial ways, or even making the universe significantly more transparent to gamma rays at certain energies.

Several studies by \cite{2009MNRAS.394L..21D, dominguez11b, 2012PhRvD..86h5036T, PhysRevD.87.035027} have reported hints of such coupling between gamma-ray photons coming from blazars and the hypothetical bosons over the past years. They are all based on observations of an apparent hardening or "pile-up"  in the estimated intrinsic VHE spectra of several blazars, once observations are corrected for the effect of the EBL according to a given model. Other authors have found no evidence of this sort of anomaly (e.g. \citealt{biteau15, 2013A&A...554A..75S, dominguez15}) .

In order to test the agreement between our results and previous studies suggesting the existence of such oscillations, we present, in Figure~\ref{Residuals}, the fit residuals (in standard deviations) as a function of the EBL optical depth (as predicted by the D11 template). Each of the points corresponds to one bin of estimated energy in one of the 32 VHE spectra of the analyzed sample. In the analysis, all the bins containing at least one on- or off-source event are used, regardless of whether or not there was a significant excess of gamma-like events from the source in the bin\footnote{The VHE SEDs obtained with MAGIC which are shown throughout the paper only show points with significant gamma-ray excesses (relative statistical uncertainty of the flux smaller than 50\%) - but the number of energy bins used in the analysis is much larger.}. This approach avoids the possible bias resulting, in the low-statistics regime, from keeping upward-fluctuating spectral points while rejecting those under the noise level. The estimated energy bins in which there is a gamma-ray excess larger than 1.5 $\sigma$ are shown in Figure~\ref{Residuals} as filled symbols, and display the expected bias towards positive values, particularly at high optical depths. When all bins in the analysis are considered, fit residuals show, at all optical depths, the expected behavior in absence of anomalies, fluctuating around 0. We have defined ten bins in $\tau$, and computed the average residuals in them. The results, both for the analysis which uses only MAGIC data and the one using MAGIC+\lat, show that the observed attenuation is  compatible with the optical depth predictions from the D11 EBL model. When the analysis is repeated using for each spectrum only the range of estimated energy within which all bins have a $\geq 1.5 \sigma$ gamma-like excess (Figure~\ref{Residuals_significancecut}), the residual for the highest optical depth bin becomes significantly biased towards positive values. In our case the effect is modest ($\simeq 3 \sigma$), but we think it may be partly responsible for the above mentioned claims of anomalous transparency found in the literature. For this reason, we have also included bins without significant gamma-ray excesses in the computation of the Likelihood from which we derive our EBL constraints.
\section{Wavelength-resolved EBL determination}\label{sec_wavelength_resolved_ebl}
\begin{figure}
   \includegraphics[width=\columnwidth]{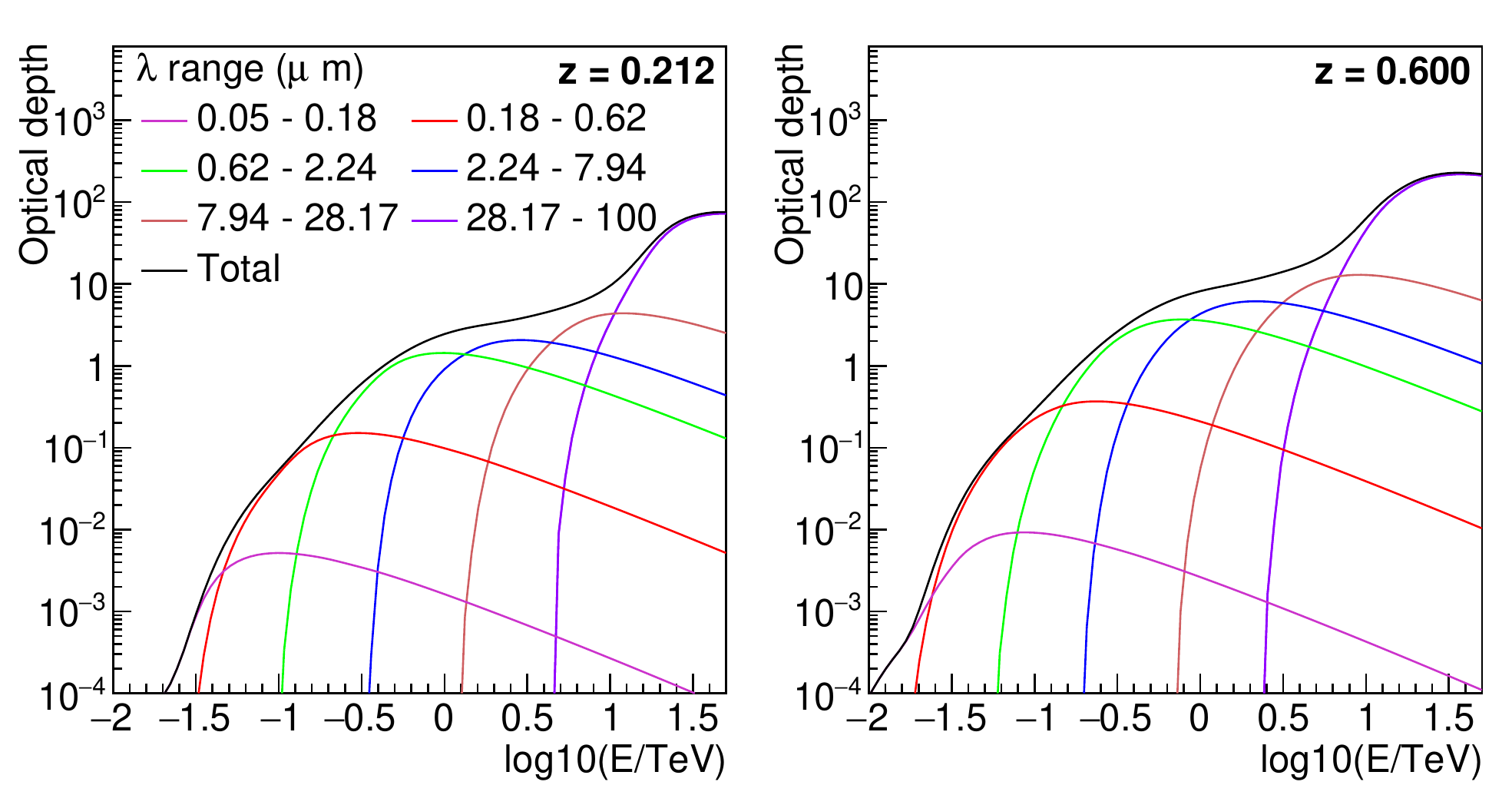}
   \vspace{-0.3cm}
   \caption{Optical depth due to the EBL in five different ranges of wavelength (units: microns), for the D11 model and two different redshifts. The optical depth in each of the five ranges is scaled by an independent free parameter during the likelihood maximization. \label{TauVsLambda}}
\end{figure}

Throughout section \ref{density_constraints} we have assumed that both the evolution and the shape of the spectrum of the EBL were exactly those of the template EBL model adopted in each case, i.e., the energy- and redshift-dependence of the total optical depth were fixed, with a single overall scaling factor as free EBL parameter. For three of the templates (D11, Fi10, G12) we have available, besides the total optical depth $\tau(E,z)$, the optical depths due to the EBL in six independent wavelength ranges, limited by the values $\lambda = $ 0.05, 0.18, 0.62, 2.24, 7.94, 28.17 and 100 ${\rm \mu m}$ (see Figure~\ref{TauVsLambda}), where the values of $\lambda$ correspond to the EBL wavelengths at $z=0$. 
\subsection{Methodology}
By scaling each of these six $\tau_i(E, z)$ with an independent factor $\alpha_i$, one can obtain the total optical depth as $\tau(E, z) = \sum_i \alpha_i \; \tau_i(E, z)$. The six values $\alpha_i$ can then be treated as independent free parameters in the likelihood maximization (see section \ref{L_maximization}), providing a handle on the shape of the EBL spectrum. Note that we do not introduce any correlation between the six parameters to impose a "smooth" EBL spectrum. The total optical depth for a given $z$, however, will behave quite smoothly  vs. $E$ because each of the  $\tau_i(E)$ curves has significant overlap with the ones of the neighboring EBL wavelength bins. The evolution of the EBL - which determines the redshift dependence of the $\tau_i$ values - is still fixed to the one of the given EBL model. Finally, the $1\,\sigma$ statistical uncertainties for the best-fit values of each $\alpha_i$ are obtained using MINOS \citep{Brun:1997pa,Hatlo:2005cj}, considering the rest of the EBL parameters $\alpha_j$, with $j\neq i$, as nuisance.
\par
The measurement of the near-UV portion of the EBL spectrum is a powerful proxy to study the star formation history of the universe and has important cosmological implications. In addition, this is one of the bands where EBL models diverge the most (see top panel of Figure~\ref{ScaledModelsSED}), and is directly accessible through observations of GRBs at high redshifts with \lat, as described by \cite{2017ApJ...850...73D}. It has to be noted that for MAGIC the optical depth $\tau_1$ due to the EBL in the first of the wavelength bins $(0.05 - 0.18 {\rm \mu m})$ is, according to the considered models, smaller than $10^{-2}$ for all energies and redshifts in our data sample (see Figure~\ref{TauVsLambda}). Therefore, its influence on the likelihood is negligible, and we cannot effectively constrain the corresponding scaling factor $\alpha_1$ (unless it was considerably larger than 1, well above the models). Besides, on Figure~\ref{TauVsLambda} it can be seen that the contribution of $\tau_1$ to the total $\tau$ is rather degenerate with that of $\tau_2$, which results in problematic $-2\log L$ minima if $\alpha_1$ is allowed to take arbitrarily large values. To address this issue, we have simply constrained $\alpha_1$ to be less than 5, meaning that the EBL density in that wavelength range is less than 5 times the EBL model estimate. The results of the fit for $\alpha_1$ typically cover the whole allowed range 0 - 5 at the $1\, \sigma_{\rm stat}$ level, and are therefore not reported since they provide no useful information.
\par
The additional freedom in the EBL modeling (relative to the simple fitting of the overall EBL density) naturally increases the degeneracy between the intrinsic spectral parameters and the EBL parameters. Consequently, the wavelength-dependent EBL determination is only possible using the MAGIC data together with the \lat\ constraints: without the latter, the method fails to converge in most of the cases on a valid minimum of the $-2\log L$ function, due to the large degeneracy between the EBL and the spectral curvature. It is important to note that this cannot be overcome by simply reducing the number of parameters of intrinsic spectral models, since it would result in the EBL model "absorbing" intrinsic features of the source spectra. For each spectrum, the same intrinsic spectral {\it model} that was chosen in the determination of the EBL density was used. The fit is however started from scratch - neither the intrinsic spectral parameters nor the EBL scaling factors from the previous single-parameter EBL density determination are known to the multi-EBL-parameter fitting algorithm. With this procedure we can test whether our blazar data {\it prefer} an EBL spectral shape different from the one in D11, Fi10 and G12, through a LRT in which the two competing models differ in the number of free EBL parameters.
\subsection{Results}
\begin{figure*}
	\includegraphics[width=2.1\columnwidth]{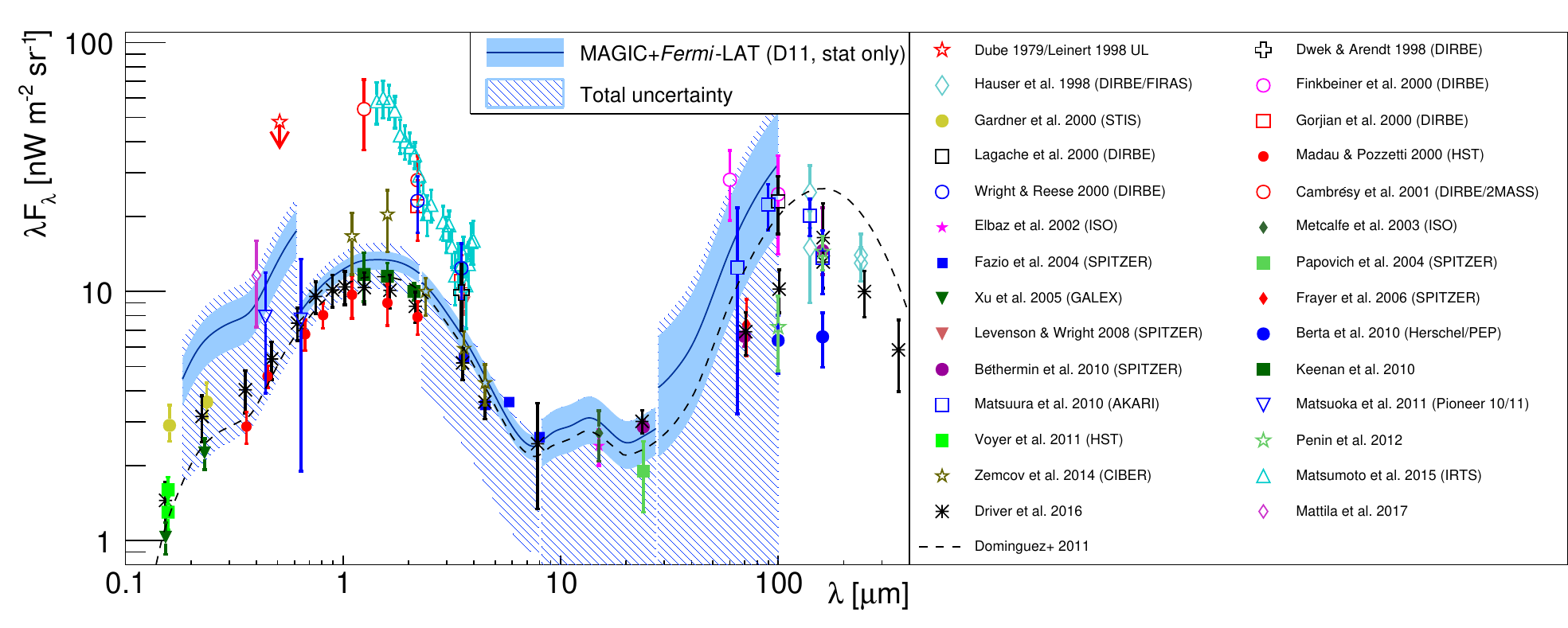}
    \caption{Wavelength-resolved EBL measurement using MAGIC and \lat\ observations, relative to the D11 EBL model at $z=0$, in five wavelength bins. A collection of direct EBL measurements is shown for comparison, taken from \citealt{2010A&A...518L..30B, 2010A&A...512A..78B, 2001ApJ...555..563C, driver16, 1979ApJ...232..333D, 1998A&AS..127....1L, 1998ApJ...508L...9D, 2002A&A...384..848E, fazio04, 2000ApJ...544...81F, 2006ApJ...647L...9F, 1538-3881-119-2-486, gorjian00, 1998ApJ...508...25H, keenan10, 2000A&A...354..247L,  2008ApJ...683..585L, madau00, mattila17, 2015ApJ...807...57M, matsuoka11, 2011ApJ...737....2M, 2003A&A...407..791M, 2004ApJS..154...70P, 2012A&A...543A.123P, 2011ApJ...736...80V, 2000ApJ...545...43W, 1538-4357-619-1-L11, zemcov14}. 
Filled symbols correspond to galaxy counts and should therefore be interpreted as lower limits.}
    \label{WaveDep_D11_vs_direct}
\end{figure*}

Using the procedure described above, we fit the data using individual optical depth scaling factors for each EBL wavelength bin. We perform the analysis independently for the three EBL templates for which we have wavelength-resolved optical depths, i.e., D11, Fi10 and G12. In each case, best-fit scaling factors, statistical uncertainties and p-values are obtained and reported in Table~\ref{tab:wavedeptable1}, and the resulting EBL SED for the D11 template is shown in Figure~\ref{WaveDep_D11_vs_direct}. As a reference, the figure also shows direct measurements (open markers), galaxy counts measurements (filled markers, to be interpreted as lower limits) and the SED of the EBL from the template of D11. The results in all bands but the one in the range of $0.18-0.62\; {\rm \mu m}$ are compatible with the template of D11 within $1\,\sigma_{\rm stat}$. Note that in the calculation of the uncertainties of a given $\alpha_i$, all the other $\alpha_j$ ($j \neq i$) are treated as nuisance parameters. For the D11 case, the fit has a total of 121 free parameters (including the 6 $\alpha_i$ factors). The number of data points is 585 (521 energy bins of MAGIC spectra + 32 \lat\ fluxes + 32 \lat\ photon indices, see appendix \ref{LikelihoodWithFermi}), hence resulting in 464 degrees of freedom. Since the intrinsic spectral models are the same as in the single-parameter EBL density determination, this fit has five additional free parameters. Comparing the final $L_{\rm max}$ values in both cases, we have $\Delta(-2\log(L_{\rm max})) = \Delta \chi^2$ = 12.37. This means, for $\Delta n_{\textnormal{d.o.f.}} = 5$, that our wavelength-resolved best-fit model is only marginally favored by the data with respect to the globally scaled D11 model in Table~\ref{tab:scalefactors}, at the $2.1 \sigma$ level\footnote{$2.4\;\sigma$ for $\Delta n_{\textnormal{d.o.f.}} = 4$, if we consider that $\alpha_1$ is constrained to be between 0 and 5, and hence not a completely free parameter.}. The situation is similar for the Fi10 model (wavelength-resolved best-fit model preferred at the $2.4\sigma$ level), whereas for the G12 model the preference for a modified EBL spectral shape relative to the one in the model is even weaker ($1.2\sigma$).

\begin{table*}
\begin{center}
\caption{Wavelength-resolved EBL constraints (scaling factors relative to three models) using MAGIC + \lat\ spectra.\label{tab:wavedeptable1}}
\setlength{\tabcolsep}{5pt}
\begin{tabular}{lrrrrrr}
\multicolumn{1}{c|}{} &   \multicolumn{5}{|c||}{EBL wavelength range (${\rm \mu m}$, @ z=0)} & p-value\\
\hline
EBL model & 0.18 - 0.62 & 0.62 - 2.24 & 2.24 - 7.94 & 7.94 - 28.17 & 28.17 - 100 & \\
\hline
D11 (stat): & 2.60 $(+0.56, -0.57)$  &  1.17 $(+0.09, -0.10)$  &   1.10 $(+0.12, -0.13)$ & 1.13 $(+0.25, -0.24)$ & 1.62 $(+0.99, -0.77)$ & $1.15\times 10^{-3}$ \\
(stat+sys): & $(+0.93, -1.72)$ & $(+0.19, -0.27)$ & $(+0.15, -0.69)$ & $(+0.25, -1.13)$ & $(+1.31, -1.62)$ & \\
\hline
Fi10 (stat): & 1.89 $(+0.58, -0.33)$ & 1.04 $(+0.12, -0.06)$ &  1.05 $(+0.11, -0.10)$ & 0.68 $(+0.18, -0.16)$ & 2.40 $(+1.65, -1.32)$ & $0.91\times 10^{-3}$ \\
(stat+sys): & $(+0.77, -1.10)$ & $(+0.25, -0.23)$ & $(+0.17, -0.59)$ & $(+0.37, -0.68)$ & $(+4.24, -2.40)$ & \\
\hline
G12 (stat): & 1.45 $(+0.35, -0.26)$ & 1.01 $(+0.10, -0.07)$ &  1.03 $(+0.10, -0.10)$ & 1.16 $(+0.27, -0.25)$ & 2.01 $(+1.13, -0.89)$ & $1.53\times 10^{-3}$ \\
(stat+sys): & $(+0.82, -0.97)$ & $(+0.25, -0.22)$ & $(+0.16, -0.63)$ & $(+0.30, -1.16)$ & $ (+1.13, -2.01)$ & \\
\hline
\end{tabular}
\end{center}
\end{table*}
\begin{table}
\begin{center}
\caption{Wavelength-resolved $\lambda F_{\lambda}$ EBL constraints from MAGIC + \lat\ spectra for three models,D11, Fi10, G12, evaluated at the center of the $\lambda$ ranges in Table~\ref{tab:wavedeptable1}. For each model, the first column shows the best-fit value, the other two are the statistical and total uncertainties respectively. Units are $(\rm nW\;\;m^{-2}\;\;sr^{-1} )$.\label{tab:wavedeptable2}}
\setlength{\tabcolsep}{4pt}
\begin{tabular}{cccccccccc}
$\lambda\;(\rm \mu m)$ & \multicolumn{3}{c}{------ D11 ------} & \multicolumn{3}{c}{------ Fi10 ------} & \multicolumn{3}{c}{------ G12 ------} \\
\hline
0.33 & 7.8 & $^{+1.6}_{-1.7}$ & $^{+2.8}_{-5.1}$ & 6.6 & $^{+2.0}_{-1.1}$  & $^{+2.7}_{-3.8}$ & 6.9 & $^{+1.7}_{-1.2}$  & $^{+3.9}_{-4.6}$\\
\hline
1.18 & 13.2 & $^{+1.0}_{-1.1}$ & $^{+2.1}_{-3.1}$ & 12.8 & $^{+1.4}_{-0.7}$  & $^{+3.1}_{-2.8}$ & 12.8 & $^{+1.3}_{-0.9}$  & $^{+3.2}_{-2.7}$\\
\hline
4.22 & 5.1 & $^{+0.6}_{-0.6}$ & $^{+0.7}_{-3.2}$ & 5.1 & $^{+0.6}_{-0.5}$  & $^{+0.8}_{-2.9}$ & 4.9 & $^{+0.5}_{-0.5}$  & $^{+0.8}_{-3.0}$\\
\hline
15.0 & 3.0 & $^{+0.7}_{-0.6}$ & $^{+0.7}_{-3.0}$ & 2.2 & $^{+0.6}_{-0.5}$  & $^{+1.2}_{-2.2}$ & 2.7 & $^{+0.6}_{-0.6}$  & $^{+0.7}_{-2.8}$\\
\hline
53.1 & 11.1 & $^{+6.8}_{-5.3}$ & $^{+9.0}_{-11.1}$ & 5.4 & $^{+3.7}_{-3.0}$  & $^{+9.5}_{-5.4}$ & 10.3 & $^{+5.8}_{-4.6}$  & $^{+5.8}_{-10.3}$\\
\hline
\end{tabular}
\end{center}
\end{table}
\par
\subsection{Systematic uncertainties}
Systematic uncertainties are evaluated with the same approach described in section \ref{systematics}. The systematic uncertainty band shown in Figure~\ref{WaveDep_D11_vs_direct} for each wavelength bin is the envelope of the $1\,\sigma_{\rm stat}$ bands for all the analyses performed (with different intrinsic spectral models selection and different instrument response functions). Note that the effect of systematic uncertainties is most relevant at the low end of the uncertainty band - which reaches 0 (no EBL) for the two wavelength bins above 7.94 ${\rm \mu m}$. Even when only the systematics associated to the choice of intrinsic spectral models are considered, the result is basically the same - meaning that, for this dataset, reducing the {\it assumed} systematic uncertainty on the average absolute calibration of the telescopes would not improve the result significantly. On the other hand, a hypothetical reduction of the actual systematic {\it error} via for example run-wise correction of the data, could well result in a reduction of the statistical uncertainties of the measurement. The main limitation of this technique is currently that the effect of the EBL on the VHE spectra is, for most of the explored wavelength range, hardly discernible from plausible intrinsic spectral features like cut-offs. Only in the 0.62--2.24 ${\rm \mu m}$ range is the lower end of the systematic uncertainty band clearly above 0, because the constraint in this range is dominated by the inflection point in the $\tau$ vs $log(E)$ curves at around 1 TeV, a feature which in our sample is most visible in the SED of 1ES 1011+496 (see Figure~\ref{SEDs_D11} and \citealt{ahnen16}). Since such a feature cannot be fitted by any of the considered intrinsic spectral models (all of which are concave functions, with no inflection points), a reduction of the EBL density from its best-fit value results in a fast worsening of the fit quality, hence providing a meaningful lower bound. On the other hand, the high end of the uncertainty bands is determined mainly by the fact that for too high EBL density, the intrinsic spectra would have to become convex (which is forbidden by construction) to reproduce the MAGIC observations. These upper constraints are only slightly increased when systematic uncertainties are taken into account.
\par
\begin{figure*}
	\includegraphics[width=2\columnwidth]{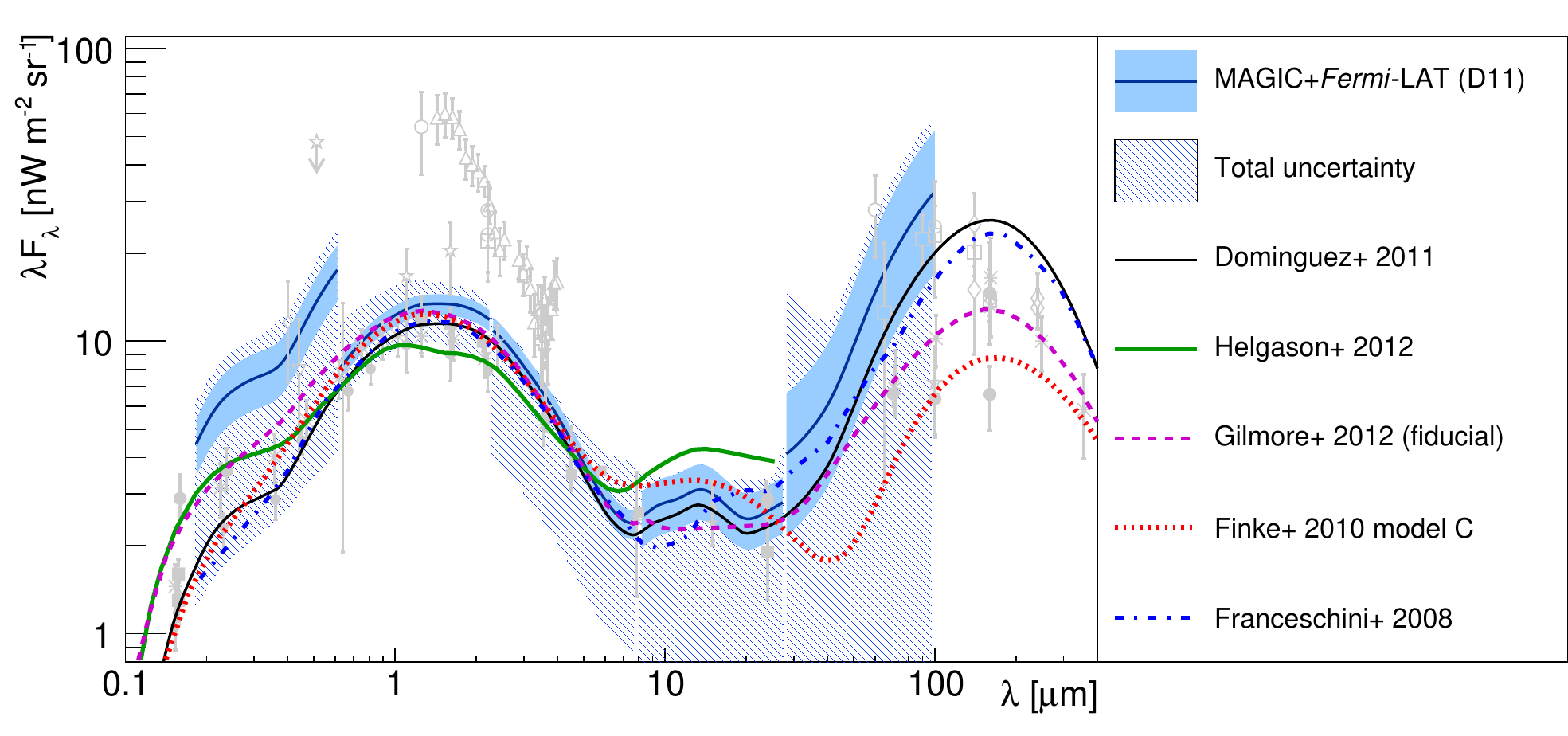}
    \caption{Wavelength-resolved EBL measurement using MAGIC and \lat\ observations. The solid band, obtained with the D11 model, is the same as in Figure~\ref{WaveDep_D11_vs_direct}; the systematic uncertainty band is the envelope of the bands obtained with three EBL templates (D11, Fi10 and G12). The result is compared to the EBL SED (at $z = 0$) for several models. Light gray symbols are the direct measurements shown in Figure~\ref{WaveDep_D11_vs_direct}.}
    \label{WaveDep_3models_vs_models}
\end{figure*}
\begin{figure*}
	\vspace{-0.5cm}
	\includegraphics[width=2\columnwidth]{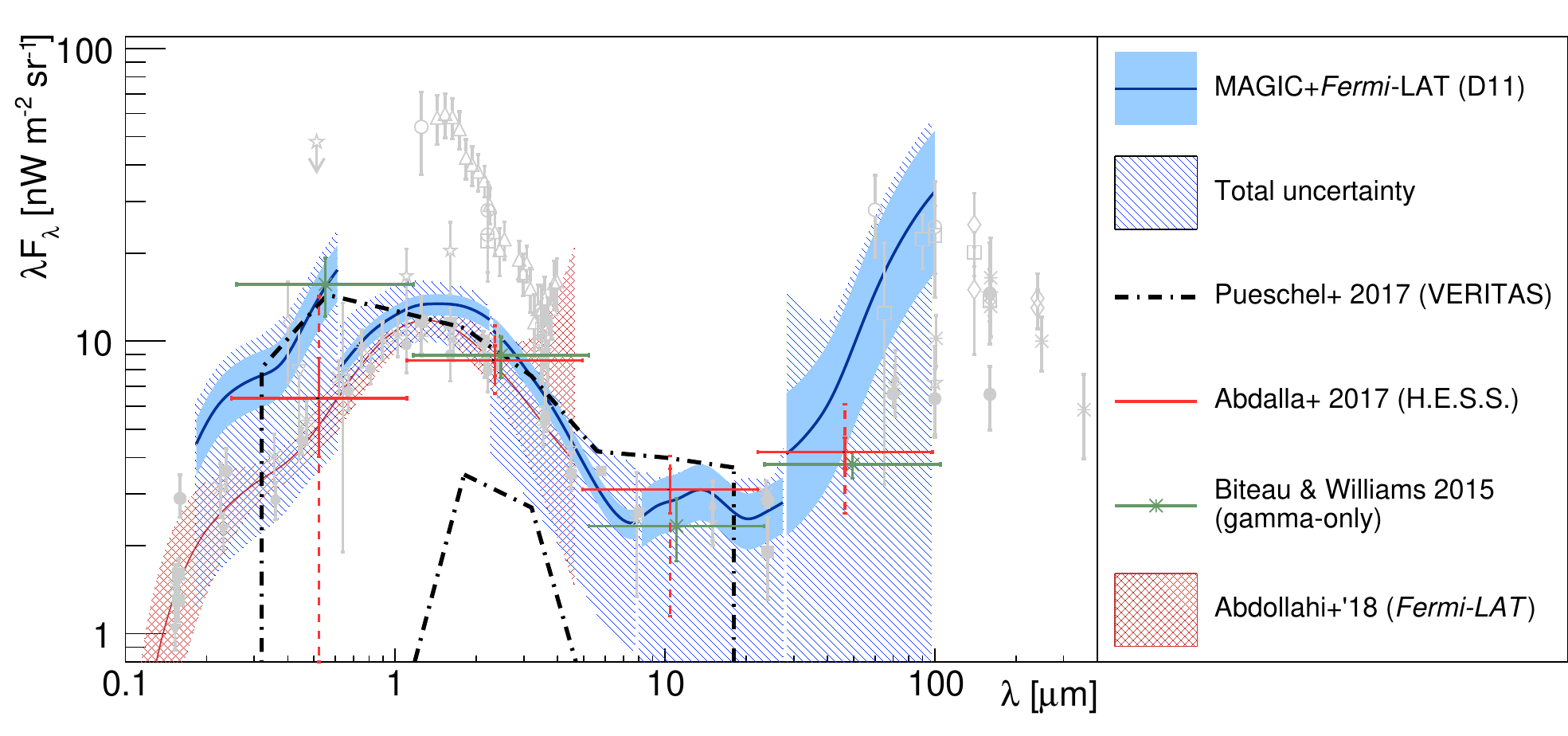}
    \caption{Wavelength-resolved EBL measurement using MAGIC and \lat\ observations (same as in Figure~\ref{WaveDep_3models_vs_models}) compared to other EBL measurements obtained with gamma-ray observations, taken from \protect\cite{Pueschel_EBL_ICRC_2017}, \protect\cite{HESS2017ebl}, \protect\cite{biteau15}, \protect\cite{FermiEBL2018}}
    \label{WaveDep_3models_vs_gamma}
\end{figure*}

We also obtained wavelength-resolved EBL measurements like those shown in Figure~\ref{WaveDep_D11_vs_direct} using the G12 and Fi10 EBL models. This allowed us to estimate the contribution of the choice of the EBL template (spectrum and evolution) to the total systematic uncertainty. The envelope of the total uncertainty bands of the three analyses (D11, G12, Fi10) is shown as the hashed area in Figure~\ref{WaveDep_3models_vs_models}, where the results are compared to five EBL models (including those used in the calculations), and again on Figure~\ref{WaveDep_3models_vs_gamma}, which displays also other measurements based on gamma-ray observations. The corresponding $\lambda F_\lambda$ values and uncertainties at the center of the wavelength bins are reported on Table~\ref{tab:wavedeptable2} for the wavelength-resolved analyses carried out with the D11, Fi10 and G12 templates. 


\subsection{Discussion}\label{discussion}

The most constraining results we obtain correspond to the two bins in the range $0.62 - 7.94\; {\rm \mu m}$, for which the statistical uncertainties in the EBL density are around 10\%. When systematic uncertainties are considered, only the range $0.62 -2.24\; {\rm \mu m}$ provides a meaningful lower bound on the EBL density, which is at the level of the galaxy counts measurements reported in \cite{madau00} (the filled red points in Figure~\ref{WaveDep_D11_vs_direct}). The best-fit values and the upper end of the total uncertainty band are respectively $\simeq 30\%$ and $\simeq 75\%$ above those measurements, meaning that most of the EBL in that wavelength range (red and near-infrared) is already resolved in individual galaxies. The result is $17\%$ ($1.9\; \sigma_{\rm stat}$) above the D11 model (which models the total EBL, including the contribution from unresolved sources of known classes), but compatible with it within systematics. In the few ${\rm \mu m}$ range, our results are clearly inconsistent with the direct measurements reported in \cite{2015ApJ...807...57M}, indicating that the large excess of isotropic near-infrared emission claimed in that work is not of extragalactic origin. At wavelengths above $7.94\; {\rm \mu m}$, where direct measurements are scarce, and EBL models differ significantly, our results are compatible with all the considered models at the $\simeq 1\;\sigma_{\rm stat}$ level, and even with zero (no EBL) within systematics.
\par
For the shortest-wavelength bin considered in this study, $0.05-0.18 \,{\rm \mu m}$ ({\it not} displayed in Figs. \ref{WaveDep_D11_vs_direct} to \ref{WaveDep_3models_vs_gamma}), the optical depths from the interaction of VHE gamma rays with such short wavelength EBL photons are simply too low for the range of redshifts (and gamma-ray energies) covered by our sample. In the $0.18 - 0.62\; {\rm \mu m}$ range our result is 2.8 $\sigma_{\rm stat}$ above the EBL density in the D11 model (2.7 and 1.7 $\sigma_{\rm stat}$ respectively for Fi10 and G12), see Table~\ref{tab:wavedeptable1} and Figure~\ref{WaveDep_D11_vs_direct}. This hint of higher-than-expected EBL in the UV-visible may well be just the result of systematic uncertainties - note that the EBL density from all three models is within the estimated systematic uncertainty band. But it is interesting to note that our result matches the direct measurement from \cite{mattila17} using the "dark cloud" method (Figure~\ref{WaveDep_D11_vs_direct}). In contrast, the estimate reported by \cite{FermiEBL2018} in the range $0.09-4.5\, {\rm \mu m}$, based on \lat\ observations of a sample of 739 blazars up to redshift $z\simeq 3.0$, is in good agreement with the D11 model. 
\par
The short-wavelength EBL excess in our analysis is strongly reduced if the five PG 1553+113 spectra are excluded from the sample. In such case, the best-fit scale factor becomes $\alpha_{ 0.18-0.62 \,{\rm \mu m}} = 1.6 \pm 0.9_{\rm stat}$, which is compatible with 1. Indeed, PG~1553+113, given its redshift of $z \geq 0.43$ and the good quality of the obtained spectra, dominates the measurement at these EBL wavelengths. Its effect on the result may seem at odds with the outcome of the MAGIC+\lat\ single-EBL-parameter analysis when only the five spectra from this source are used. In that case, the best-fit scale was well below 1 (see the corresponding profile likelihood curve in the bottom panel of Figure~\ref{ProfileLikelihood}). We must remark, however, that in the adopted method, the best-fit EBL values are those which result in the maximum likelihood with plausible {\it shapes} of the intrinsic spectra. The absolute fluxes are irrelevant, as we do not measure absolute absorption factors (because the intrinsic spectra are not known). Therefore, a given observed VHE spectrum is not bound to shift the EBL results always in a given direction, but its effect depends on the other spectra included in the sample. In the case at hand, the rest of the sources constrain the EBL in the 0.62-2.24 $\mu m$ to be close to the one in the models, and hence a low EBL in the $<0.62\, {\rm \mu m}$ range would probably imply an unnatural (significantly convex) intrinsic spectrum for PG~1553+113. This just shows that there is {\it no contradiction} in the different effect of PG~1553+113 on the two types of analysis - but provides no insight on whether the hint of an excess is genuine or not.
\par
A comparison of our results with previous EBL constraints based on gamma-ray observations is shown in Figure~\ref{WaveDep_3models_vs_gamma}. Those which include an evaluation of systematic uncertainties \citep{HESS2017ebl,Pueschel_EBL_ICRC_2017} show similar features to our own, with weak or no lower constraint except in the few ${\rm \mu m}$ range, and similar upper bounds.
\vspace{-0.5cm}
\section{Summary and conclusions}\label{summaryandconclusions}
We presented a measurement of the EBL using MAGIC and \lat\ gamma-ray observations of 12 blazars in different periods, for a total of 32 spectra. A model of the EBL (with one or more free parameters) and a set of plausible models for the 32 intrinsic spectra are used to construct a likelihood. This likelihood is then maximized to obtain the EBL model parameters most compatible with the MAGIC and \lat\ observations. The main results are the following:
\begin{itemize}
\item With only one free EBL parameter (global EBL density for fixed SED and evolution) it is possible to set constraints both with the MAGIC data alone, and with the combination of MAGIC and \lat\ contemporaneous observations. The results, shown in tables \ref{tab:scalefactors} and \ref{tab:systematicstable}, are compatible at the 1~$\sigma_{\rm stat}$ level with the EBL density in the D11, Fi10, F08 and G12 templates. The other four tested templates (H12, I13, S16 and K10) are in worse agreement with our observations, with EBL densities up to 3.5 $\sigma_{\rm stat}$ off the best-fit values. The data do not allow to discriminate clearly among the models, although the first four seem favored by our results. 

\item An assessment of the total uncertainties including systematics was performed by repeating the analysis in different conditions to account for the uncertain knowledge of the intrinsic spectral shapes and the absolute calibration of MAGIC. For the favored models, the resulting upper bounds are between 13\% and 23\% above the EBL densities in the models. We conclude that our result and the described methodology are currently dominated by systematic uncertainties.

\item The distribution of fit residuals as a function of optical depth for the D11 template (Figure~\ref{Residuals}) shows no hint of significant deviations. Therefore, we find no evidence of anomalies in the transparency of the universe to gamma rays, like those that might be attributed e.g. to photon-ALP conversions. We think that the inclusion in the analysis of bins of estimated energy for which no significant excess (or even a deficit) of gamma-like events is recorded is instrumental to avoid biases which could be misinterpreted as anomalies in the gamma-ray absorption process.
\item For the D11 EBL template we repeated the analysis by dividing the sample in four bins of redshift (in the range of $z$ from 0 to 1), with the aim of probing the evolution of the EBL density relative to that in the model. The measured constraints are compatible with the model in all four bins at the $\simeq 1\,\sigma_{\rm stat+sys}$ level (Figure \ref{Scale_vs_z_D11}). For redshifts above 0.3 the data do not allow to set any lower bound to the EBL density, once systematic uncertainties are taken into account. This is due to the degeneracy between possible intrinsic features of VHE spectra, like cut-offs, and the effect of the EBL in the range of energies accessible by MAGIC at such distances. The measured optical depth for the last redshift bin is however significantly above $\alpha=0$ for the MAGIC+\lat\ analysis if only statistical uncertainties are considered. It underlines the fact that, for the first time in the VHE band, we are able to effectively explore the EBL at redshift close to 1. It also sets promising prospects for future instrumentation with reduced systematic uncertainties.

\item The combination of MAGIC and \lat\ data allows to perform $\lambda$-resolved measurements of the EBL, for three of the models (D11, Fi10, G12), for which we have available the optical depths as a function of EBL wavelength (tables \ref{tab:wavedeptable1} and \ref{tab:wavedeptable2} and figs. \ref{WaveDep_D11_vs_direct} to \ref{WaveDep_3models_vs_gamma}). This procedure shows that the constraints obtained with our blazar data sample are mostly driven by the EBL in the $\lambda \simeq 0.6$ to $8 \; {\rm \mu m}$ range. The upper EBL bound in that range, including systematic uncertainties, is between 13 and 29\% higher than the models, leaving little room for additional EBL contributions not accounted for in the models. In the 0.18 - 0.62 ${\rm \mu m}$ range we obtain a relatively high EBL density, particularly with respect to the D11 and Fi10 models, but the deviation is not significant once systematic uncertainties are considered.
\end{itemize}

\par
Finally, it must be stressed once more that the method used in the present paper (and in previous similar works in the literature) for the determination of the EBL relies on the assumption that the chosen spectral blazar models can reproduce the intrinsic spectra of the blazars in the sample. We adopted a conservative approach by always allowing the intrinsic spectra to be curved (pure power laws were only tried for the estimation of systematic uncertainties). In addition, we just required a better p-value in order to adopt a more complex model (e.g. a log parabola with exponential cut-off over a log parabola), instead of a minimum significance of the corresponding LRT (e.g. 2 $\sigma$ in \citealt{biteau15}). These differences in the methods to select intrinsic spectral models make it impossible to compare directly the merit of the different gamma-ray based EBL measurements shown in Figure~\ref{WaveDep_3models_vs_gamma}. 
\par
We expect that with a large sample of high-quality spectra up to the few TeV range, such as those that will be obtained in the coming years with the Cherenkov Telescope Array (CTA, \citealt{acharya13}), it will be possible to relax the assumption on the intrinsic spectral shapes, to include more general concave functions beyond those used in this work. CTA will also benefit from a better control of the systematics related with the atmospheric conditions and the absolute calibration of the telescopes. Together with the increased redshift range provided by its lower energy threshold relative to current IACTs, CTA will certainly be a major contributor to the study of the EBL.
\par
\vspace{-0.5cm}
\section*{Acknowledgements}
We would like to thank the Instituto de Astrof\'{\i}sica de Canarias for the excellent working conditions at the Observatorio del Roque de los Muchachos in La Palma. The financial support of the German BMBF and MPG, the Italian INFN and INAF, the Swiss National Fund SNF, the ERDF under the Spanish MINECO (FPA2015-69818-P, FPA2012-36668, FPA2015-68378-P, FPA2015-69210-C6-2-R, FPA2015-69210-C6-4-R, FPA2015-69210-C6-6-R, AYA2015-71042-P, AYA2016-76012-C3-1-P, ESP2015-71662-C2-2-P, FPA2017‐90566‐REDC), the Indian Department of Atomic Energy, the Japanese JSPS and MEXT and the Bulgarian Ministry of Education and Science, National RI Roadmap Project DO1-153/28.08.2018 is gratefully acknowledged. This work was also supported by the Spanish Centro de Excelencia ``Severo Ochoa'' SEV-2016-0588 and SEV-2015-0548, and Unidad de Excelencia ``Mar\'{\i}a de Maeztu'' MDM-2014-0369, by the Croatian Science Foundation (HrZZ) Project IP-2016-06-9782 and the University of Rijeka Project 13.12.1.3.02, by the DFG Collaborative Research Centers SFB823/C4 and SFB876/C3, the Polish National Research Centre grant UMO-2016/22/M/ST9/00382 and by the Brazilian MCTIC, CNPq and FAPERJ. The work of the author M. Nievas Rosillo is financed with grant FPU13/00618 of MECD. A. Dom{\'i}nguez thanks the support of the Ram{\'o}n y Cajal program from the Spanish MINECO. The work of the author M. V\'azquez Acosta is financed with grant RYC-2013-14660 of MINECO.



\vspace{-0.5cm}
\bibliographystyle{mnras}
\bibliography{biblio} 



\appendix

\section{Maximum likelihood method}
\label{LikelihoodMaxim}
In this analysis we have used a joint (i.e., many-spectra) maximum likelihood approach, similar to those used in \cite{lat_ebl10, ackermann12, abramowski13,FermiEBL2018}. The method is implemented in the ROOT-based \cite{Brun:1997pa} MARS software package \cite{moralejo09, zaninMARS, MAGICperformance2016}, which is the official analysis of the MAGIC collaboration. The joint likelihood $L$ to be maximized is the product of a number of factors, one for every bin ($j$) in estimated energy of every gamma-ray spectrum ($i$) used in the analysis:
\begin{equation}
L (ebl, \theta_1, \theta_2, ..., \theta_{N_{\text{spectra}}}, b) = \displaystyle\prod_{i=1}^{N_{\text{spectra}}}\displaystyle\prod_{j=1}^{N_{\text{bins,i}}} L_{ij}(ebl, \theta_i, b_{ij})
\end{equation}
where each $\theta_i$ is a vector containing the parameters describing the {\it intrinsic} spectrum $i$ (which are treated as nuisance parameters in the likelihood maximization), and $ebl$ is a vector of parameters (or a single parameter) describing the EBL. The parameters $b_{ij}$ are nuisance parameters related to the poissonian background recorded together with the gamma-ray signal. Each factor $L_{ij}$ has the form:
\begin{equation}
\begin{split}
L_{ij} (ebl, \theta_i) = \; & Poisson(g_{ij}(ebl, \theta_i)+b_{ij}, \; N_{\text{on},ij})\;\cdot \\
& Poisson(b_{ij}/ \beta, \; N_{\text{off},ij})
\end{split}
\label{Likelihoodij}
\end{equation}
Here $N_{on}$ and $N_{off}$ are the numbers of recorded events (after gamma-ray selection cuts) in bins of estimated energy ($j = 1,..., N_{\text{bins},\;i}$), both around the source direction (ON-source region), $N_{\text{on},ij}$, and in three control regions of identical size (OFF) which contain only background events, $N_{\text{off},ij}$. The Poisson parameters for the signal and the background are respectively $g_{ij}$ and $b_{ij}$, which are described in more detail in the next paragraph. The factor $\beta$ is the ratio of ON to OFF exposure, which could be different for each spectrum, but happens to be the same, $\beta = 1/3$, in the analysis presented here - the three OFF sky regions considered for each observation are chosen to have the same acceptance as the ON region. In each spectrum, the $j$ index runs over bins in the range 60 GeV - 15 TeV of estimated energy. It is not required that a bin has a significant excess of gamma-like events, but the range is clipped on both ends so that all bins within it contain at least one event in the ON-source region or in the OFF-source region. This results in different fitting ranges for each observation, depending mostly on the range of zenith distance of the observations. In the present analysis of 32 spectra, the total number of considered energy bins is $\sum_{i=1}^{N_{\text{spectra}}} N_{bins, i} = 521$, and the total number of parameters needed to describe the 32 intrinsic spectra varies from 103 to 106, depending on the template EBL model used. 
\par
We follow the profile likelihood method described in \cite{Rolke:2004mj}, with the parameter(s) of interest being in our case those which describe the EBL. Each of the $L_{ij}$ terms defined in eq.~(\ref{Likelihoodij}) is the product of two poissonian probabilities: the probability of observing $N_{\text{on},ij}$ events in the ON region, and the probability of observing $N_{\text{off},ij}$ events in the OFF region. The value $g_{ij}$ is the Poisson parameter (mean number) of gammas in the ON-source region for bin $j$ of spectrum $i$, and is obtained by folding the intrinsic source spectrum (given by $\theta_i$) with the EBL absorption (according to the $ebl$ parameters), and with the MAGIC response (energy-dependent effective area and energy migration matrix), and multiplying the resulting gamma-ray rate by the observation time. The Poisson parameter of the background in the ON-source region is $b_{ij}$, and it is treated as a nuisance parameter: in each step of the likelihood maximization we look for the value of $b_{ij}$ which maximizes $L_{ij}$, given $g_{ij}$, $\beta$, $N_{\text{on},ij}$ and $N_{\text{off},ij}$. As shown by \cite{Rolke:2004mj}, the $b_{ij}$ values can be calculated analytically from the rest of the parameters and the data inputs by solving a quadratic equation.
\subsection{Treatment of statistical uncertainties in the MAGIC response}
An additional complication arises from the fact that the instrument response function of the telescopes (with which the gamma-ray spectrum has to be folded) is actually not known with perfect accuracy, since it is obtained from a Monte Carlo simulation  with limited statistics. Therefore, for given parameters $(\theta, ebl)$, the result of the folding process is not a single value $g_{ij}$ for a given bin, but rather a {\it range of values}, $g_{ij} \pm \Delta g_{ij}$. In order to account for this, we replace $g_{ij}$ in the expressions above by another nuisance parameter, $g^\prime_{ij}$. Then, assuming that the uncertainty $\Delta g_{ij}$ is gaussian (this should be the case except in case of very low MC statistics), we add another factor to the likelihood, i.e.:
\begin{equation}
\begin{split}
L_{ij} = \; & Poisson(g^\prime_{ij}+b_{ij}, \; N_{\text{on},ij})\;\cdot Poisson(b_{ij}/ \beta, \; N_{\text{off},ij}) \;\cdot \\
& Gauss(g_{ij}^\prime;\; g_{ij}, \Delta g_{ij})\; , \\
\text{with} \\
Gauss&(g^\prime_{ij}; g_{ij}, \Delta g_{ij}) = \frac{1}{\sqrt{2\pi} \; \Delta g_{ij}} e^{-\frac{1}{2}(g^\prime_{ij}-g_{ij})^2/\Delta g^2_{ij}} \\
\end{split}
\label{Likelihoodij_2}
\end{equation}
where $g_{ij}$ and $\Delta g_{ij}$ depend, as in equation~(\ref{Likelihoodij}), on $ebl$ and $\theta_i$. The gaussian factor above penalizes values of $g^\prime_{ij}$ which are too far from the MC-estimated value $g_{ij}$. This scenario (gaussian uncertainty in the detector efficiency and poissonian background) is mentioned in \cite{Rolke:2004mj}, but not explained in detail. The idea is that now, instead of maximizing each of the $L_{ij}$ terms with respect to $b_{ij}$ alone, we have to look for the values $(b_{ij}, g^\prime_{ij})$ that maximize $L_{ij}$ given $g_{ij}$, $\Delta g_{ij}$, $N_\text{on,ij}$, $N_\text{off,ij}$ and $\beta$. It turns out that (dropping the $ij$ indices for clarity), if we fix all other values, the optimal $b$ and $g^\prime$ can also be found analytically, in the general case, by solving  a third-degree equation. For the particular cases of $N_\text{on}=0$ or $N_\text{off}=0$, the solution is even simpler, and involves solving a linear and a second-degree equation respectively - always taking care of forcing $b = 0$ (or $g^\prime = 0$) in the rare cases in which the analytical solution is unphysical, i.e. $b<0$ ($g^\prime < 0$).
\subsection{Use of \lat\ constraints \label{LikelihoodWithFermi}}
Constraints from contemporaneous \lat\ spectra can be incorporated into the method by adding for each spectrum, two additional factors to the likelihood, which correspond to the comparison of the flux and photon index measured at the decorrelation energy of the LAT spectrum, $F_{\text{\tiny LAT}}\pm\Delta F_{\text{\tiny LAT}}$, $\Gamma_{\text{\tiny LAT}}\pm\Delta\Gamma_{\text{\tiny LAT}}$ (a "spectral bow-tie"), and those of the tested spectral function ($F$, $\Gamma$) at the same energy. For these terms we assume the LAT parameter uncertainties to be gaussian. 
\begin{equation}
L_{ij} = L_{ij,\text{\tiny MAGIC}} \cdot e^{-\frac{1}{2} \left(\frac{\Gamma - \Gamma_{\text{\tiny LAT}}}{\Delta \Gamma_{\text{\tiny LAT}}}\right)^2} \cdot e^{-\frac{1}{2} \left(\frac{F - F_{\text{\tiny LAT}}}{\Delta F_{\text{\tiny LAT}}}\right)^2}
\label{Likelihoodij_3}
\end{equation}
where $L_{ij,\text{\tiny MAGIC}}$ is given by expression (\ref{Likelihoodij_2}).
A 10\% systematic uncertainty in the \lat\ collection area\footnote{\protect\url{https://fermi.gsfc.nasa.gov/ssc/data/analysis/LAT_caveats.html}} has been added quadratically to the $F_{\text{\tiny LAT}}$ values. With the procedure outlined above, each of the \lat\  spectra contributes two additional {\it data points} (and degrees of freedom) to the fit. A possible improvement over this simplified approach could be achieved through the inclusion in the joint Likelihood of the contributions from each of the \lat\ spectral points. It must be remarked however that the points are correlated, and often suffer from low photon statistics, so a rigorous treatment is far from trivial.
\subsection{Sources at uncertain redshift}
When the redshift of a source is uncertain (which is only the case, in our sample, for PG 1553+113), $z$ is treated also as a nuisance parameter, with flat distribution in the allowed range. This is done by scanning the redshift, in each step of the likelihood maximization process, to maximize the contribution to the joint likelihood of the corresponding spectra.
\writeaffiliations

\bsp	
\label{lastpage}

\end{document}